\documentstyle[12pt,psfig]{article}

\newcommand{\e}{{\rm  e}}
\newcommand{\fb}{{\rm  fb}}

\newcommand{\beq}{ \begin{eqnarray} }
\newcommand{\eeq}{ \end{eqnarray} }
\newcommand{\beqstar}{ \begin{eqnarray*} }
\newcommand{\eeqstar}{ \end{eqnarray*} }
\newcommand{\gsim}{ \mathop{}_{\textstyle \sim}^{\textstyle >} }
\newcommand{\lsim}{ \mathop{}_{\textstyle \sim}^{\textstyle <} }

\newcommand{\GEV}{ {\rm GeV} }

\newcommand{\sla}[1]{\not\!#1}
\newcommand{\E}{ \sla{E} }

\newcommand{\cha}{\tilde{\chi}^+}
\newcommand{\chb}{\tilde{\chi}^-}
\newcommand{\neu}{\tilde{\chi}^0}
\newcommand{\sml}{\tilde{\mu}_L}
\newcommand{\stl}{\tilde{\tau}_L}
\newcommand{\sel}{\tilde{\e}_L}

\newcommand{\sne}{\tilde{\nu}_\e}
\newcommand{\snm}{\tilde{\nu}_\mu}
\newcommand{\snt}{\tilde{\nu}_\tau}
\newcommand{\tl}{\tilde{l}}

\newcommand{\tnu}{\tilde{\nu}}
\newcommand{\tchi}{\tilde{\chi}}

\begin{document}
\baselineskip 0.7cm

\begin{titlepage}

\begin{center}

\hfill KEK-TH-586\\
\hfill OU-HET-303\\
\hfill YITP-98-57 \\
\hfill hep-ph/9808410\\
\hfill \today

  {\large  Lepton-Flavor Violation \\
        in the Left-handed Slepton Production \\
        at Future Lepton Colliders}
  \vskip 0.5in {\large
    Junji~ Hisano$^{(a)}$,
    Mihoko M.~Nojiri$^{(b)}$,
    Yasuhiro~Shimizu$^{(a)}$, and 
    Minoru~Tanaka$^{(c)}$ }
\vskip 0.4cm 
{\it 
(a) Theory Group, KEK, Oho 1-1, Tsukuba, Ibaraki 305-0801, Japan
}
\\
{\it 
(b) YITP, Kyoto University, Kyoto 606-8502, Japan
}
\\
{\it 
(c) Department of Physics, Graduate School of Science, 
        Osaka University, Toyonaka, Osaka 560-0043, Japan
}
\vskip 0.5in

\abstract {
The Super-Kamiokande atmospheric neutrino data 
suggest existence of the large lepton-flavor violating (LFV) interaction 
in the higher energy scale. If the minimal supersymmetric standard
model is extended to have right-handed neutrinos, 
the left-handed sleptons in the second and third 
generations are expected to have the LFV masses in the minimal 
supergravity scenario. In this article we study the LFV signals in the 
left-handed slepton production at $\mu^+\mu^-$ colliders
and $\e^+\e^-$ linear colliders (LC's),
$\mu^+\mu^-(\e^+\e^-)\rightarrow\tau\mu +4jets + \E$ and 
$\mu^+\mu^-(\e^+\e^-)\rightarrow\tau\mu l+ 2jets+ \E$.
The main background comes from decay of a tau lepton into a muon in 
the lepton-flavor conserving  slepton pair production.
They are significantly reduced by the energy and the impact parameter cuts
for the muon. At $\mu^+\mu^-$ colliders (LC's) it may be possible 
to reach the mixing angle $\sin 2\theta_{\tilde{\nu}} \gsim 0.1(0.5)$ and 
the mass difference $\Delta m_{\tilde{\nu}}\gsim 0.1(0.4)$ GeV for the 
sneutrinos in the second and third generations 
at the statistical significance of $3\sigma$.
Such small mass difference and a mixing angle may be induced radiatively 
even if the Yukawa coupling constant for the tau neutrino is of the 
order of 0.1.}
\end{center}
\end{titlepage}
\setcounter{footnote}{0}

\section{Introduction}

The Super-Kamiokande collaboration has reported evidence that the 
atmospheric neutrino anomaly is indeed due to neutrino oscillation
\cite{superkamiokande}. This phenomenon is lepton-flavor violating
(LFV), and it is the first convincing signature beyond the 
standard model (SM).
From the zenith-angle dependence of  $\nu_\e$ and 
$\nu_{\mu}$ fluxes following neutrino mass square difference and a mixing angle 
are favored,
\begin{eqnarray}
&\Delta m^2_{\nu_{\mu} \nu_X} \simeq (5\times10^{-4} - 6\times10^{-3}) {\rm eV^2}, &
        \nonumber\\
&\sin^2 2\theta_{\nu_{\mu} \nu_X}  >0.82.&
\end{eqnarray}
Provided mass hierarchy $m_{\nu_{\tau}}\gg m_{\nu_{\mu}}\gg m_{\nu_\e}$
it is natural to consider $\nu_X = \nu_{\tau}$ while the solar neutrino 
deficit is explained by the MSW \cite{MSW} or the vacuum oscillation
between $\nu_{\mu}$  
and $\nu_\e$.
This is also consistent with the result of 
the CHOOZ experiment \cite{chooz}.
Thus, the atmospheric neutrino anomaly implies a  
non-vanishing neutrino mass,
\begin{equation}
      m_{\nu_{\tau}} \simeq (0.02-0.08)  {\rm eV}.
\label{taumass}
\end{equation}

The simplest model to generate the neutrino masses is the seesaw 
mechanism \cite{seesaw}.
The neutrino mass Eq.~(\ref{taumass}) leads to the right-handed neutrino masses below  
$\sim$ $(10^{14}-10^{15})$ GeV, even if  the Yukawa coupling constant
of the right-handed neutrino is  of the order of 1. 
This means that a  
LFV interaction between the second and third generations exists below 
the gravitational scale ($M_G\sim 10^{18}\GEV$). 

The supersymmetric (SUSY) extension of the standard model (MSSM) is also one
of the most promising model beyond the SM \cite{nilles}. 
It is one of the solutions 
to  the naturalness problem for the Higgs mass.
The SUSY breaking masses for sleptons in the MSSM
are sensitive to the physics beyond the MSSM if they are originated
from the hidden sector in the minimal supergravity (that is,
the minimal supergravity scenario).
In particular, if the LFV interaction 
exists below the gravitational scale, the radiative correction 
generates the LFV masses for sleptons, and this 
predicts the LFV processes at low energy, $\mu\rightarrow\e\gamma$,
$\tau\rightarrow \mu\gamma$, and so on \cite{HKR}. Many 
studies have been done for them 
\cite{su5}\cite{so10}\cite{GM}\cite{BM}\cite{HMTYY}\cite{HMTY}. 
The large mixing between $\nu_\mu$ and $\nu_\tau$ suggests 
the large mixing between  the left-handed sleptons
in the second and third generations
($\tilde{\mu}_L$ and $\tilde{\tau}_L$, and $\tilde{\nu}_\mu$ and 
$\tilde{\nu}_\tau$) in the minimal supergravity scenario. One of the 
predictions is $\tau\rightarrow \mu\gamma$. However, the future 
experiments may not reach to the expected branching ratio due to the 
low sensitivity \cite{HMTY}. 

In this article, we study an alternative way, search for the LFV in the 
left-handed slepton production
at future lepton colliders, such as the  $\e^+\e^-$ linear 
colliders (LC's) and $\mu^+\mu^-$ colliders. When the left-handed sleptons 
in the second and third generations have the LFV masses, the signatures are
following,
\begin{eqnarray} 
1)~\e^+\e^-(\mu^+\mu^-)&\rightarrow&{\tilde{\nu}}{\tilde{\nu}}^c~{\rm or}~
        {\tilde{l}^+}{\tilde{l}^-}
   \rightarrow\tau\mu +4jets + \E,
\nonumber\\
2)~\e^+\e^-(\mu^+\mu^-)&\rightarrow&{\tilde{\nu}}{\tilde{\nu}}^c
\rightarrow\tau \mu l+ 2jets+ \E,
\nonumber\\
3)~\e^+\e^-(\mu^+\mu^-)&\rightarrow&{\tilde{l}^+}{\tilde{l}^-}
\rightarrow
\tau\mu ll +2jets + \E,
\label{signals}
\end{eqnarray}
with $(l=\e,\mu,\tau)$. The jets and additional leptons in the 
final states come from 
the decay of the wino-like chargino and neutralino, into which the 
left-handed sleptons decay dominantly if the decay 
modes are open.  We find that for 
the one year of  the proposed LC ($\mu^+\mu^-$ collider) run,  it may be possible 
to reach the mixing angle $\sin 2\theta_{\tilde{\nu}} \gsim 0.5(0.1)$ and 
the mass difference $\Delta m_{\tilde{\nu}}\gsim 0.4(0.1)$GeV for the 
sneutrinos in the second and third generations at the statistical significance
of $3\sigma$. 
Such  small mass difference and a mixing angle may be generated radiatively 
even if the Yukawa coupling constant of the tau neutrino is almost as large 
as that of  the tau lepton at low $\tan\beta$ ($f_{\nu_\tau}\sim 0.1$).
We may observe the LFV processes at future colliders 
if the large mixing of neutrinos originates from the large mixing of 
the neutrino Yukawa coupling constants.

Searching for these signals has some advantages.
First, these processes are at tree level while $\tau\rightarrow \mu
\gamma$ is a one-loop process. 
This means that we do not need so high statistics compared with 
$\tau\rightarrow \mu\gamma$. Also, even if the sleptons 
are degenerate in the masses, the cross sections for the 
signals are suppressed by at most 
$\Delta m_{\tilde{l}}/\Gamma_{\tilde{l}}$ with 
the mass difference of the sleptons $\Delta m_{\tilde{l}}$  
and the total width $\Gamma_{\tilde{l}}$ 
\cite{ACFH1}\cite{ACFH2}. This dependence comes from the interference 
among the real slepton  productions in the above signals. 
The decay widths for the left-handed sleptons 
are about 1GeV. On the other hand, $\tau\rightarrow\mu\gamma$ 
is strongly suppressed by $\Delta m_{\tilde{l}}/\bar{m}_{\tilde{l}}$ 
where  $\bar{m}_{\tilde{l}}$ is the average of the slepton  masses, since 
only the virtual slepton exchange contributes to it.

Second, these signals are almost free from the SM 
backgrounds (BG's) since they have two or more jets and two or more leptons 
with different flavors in the final states due to the cascade decays of the 
SUSY particles. It is shown in Ref.~\cite{BMT} that 
for the electron sneutrino production 
at LC's, $\e\e\mu +2jets+\E$ is BG free.
In the previous works 
\cite{ACFH1}\cite{ACFH2}\cite{Krasnikov}\cite{cheng}\cite{HT}
only the LFV signal in the slepton
production sequel to decay into the LSP directly is studied. 

Third, in $\mu^+\mu^-$ colliders the cross section for muon sneutrino 
production reaches to 1pb, and we may get statistics enough for search
and study of the LFV in the sneutrino masses.
The sneutrino mass matrix is determined by only the SUSY breaking mass
parameters for the left-handed sleptons, while that for 
the charged sleptons depends on them and the other parameters. Then, we 
can extract information on the SUSY breaking mass parameters for the
left-handed sleptons directly from the signal through the sneutrino production,
$\mu^+\mu^-\rightarrow\tau \mu l + 2jets+ \E$. 

Contents of this article are following.  In the next section, we discuss
the origin of the large mixing suggested from the Super-Kamiokande result,
and review the radiative generation of the LFV masses for the left-handed 
sleptons in the seesaw mechanism. In Section 3 the formula of the cross 
sections for the LFV signals in a case where the left-handed sleptons
have the LFV masses, and the signal cross sections 
are given for a representative parameter set. In 
Section 4 we discuss the BG's for the signals. The main BG's come from 
lepton-flavor conserving $\tilde{\nu}_\tau$ or $\tilde{\tau}_L$ pair
production, since $\tau$ can decay into $\mu$.
We show that the cuts on 
the energy and the impact parameter (IP)  are useful 
to separate the 
primary muon from the slepton decay and the 
secondary muon from the tau decay. Also, we show
the significance of the signals over the BG's in LC and muon
collider experiments. Section 5 is Conclusion and Discussion.
Appendices A and B
are for  our convention for the MSSM Lagrangian. In Appendix C 
we show the matrix elements for the slepton production and 
the decay. In Appendix D the three-body decay widths for the charginos and 
neutralinos are given. In Appendix E we present the formula of the energy and 
IP distribution for muon from the tau decay.

\section{Radiative generation of the LFV slepton masses in the seesaw mechanism}

In this section, we will discuss the origin of the large mixing angle for the atmospheric
neutrino observation, and explain the radiative generation of the 
LFV masses for the left-handed sleptons. Assuming that the atmospheric neutrino 
observation comes from the 
oscillation between $\nu_\mu$ and $\nu_\tau$,
the Super-Kamiokande result for the atmospheric neutrino
may mean existence of the large mixing between the second and  third 
generations in the Yukawa coupling for the neutrinos.
In this case, the mixing generates 
the LFV masses in 
$\tilde{\tau}_L$-$\tilde{\mu}_L$ and 
$\tilde{\nu}_\tau$-$\tilde{\nu}_\mu$ 
radiatively in the minimal supergravity scenario,  

The MSSM with the right-handed neutrinos (MSSMR) is 
the simplest supersymmetric model to explain the neutrino masses.  The 
superpotential of the lepton sector is given as 
\begin{eqnarray}
W_{\rm MSSM+\nu_R}&=& 
   f_{\nu_i} H_2 N^c_i L_i 
+  f_{l_i} U^\dagger_{Dij} H_1 E^c_i L_j
+  \frac12 M_{ij} N^c_i N^c_j,
\label{yukawacoupling}
\end{eqnarray}
where $L$ is left-handed leptons, and  $N^c$ and 
$E^c$ are right-handed neutrinos and charged 
leptons.\footnote{
We represent superfields in capital letters, and the components
in small letters.}
$H_1$ and $H_2$ are the Higgs doublets in the MSSM. Here, $i$ and $j$
are generation indices.
A unitary matrix $U_D$ is  similar to the Cabibbo-Kobayashi-Maskawa (CKM)
matrix in the quark sector.
After the  right-handed neutrinos are integrated out,
the mass matrix for the left-handed neutrinos is given as 
\begin{eqnarray}
(m_{\nu_L})_{ij} &=&m_{\nu_iD} \left[M^{-1}\right]_{ij} m_{\nu_jD}
\nonumber\\
&\equiv& U_{Mik}^T m_{\nu_k}  U_{Mkj},
\end{eqnarray}
where $U_M$ is a unitary matrix and 
\begin{eqnarray}
m_{\nu_i D} &=& f_{\nu_i } v \sin\beta/\sqrt{2},
\end{eqnarray}
where $\langle H_2 \rangle=(0,v\sin\beta/\sqrt{2})^T$ with $v\simeq 246$GeV.
We assume  $m_{\nu_\tau} \gg m_{\nu_\mu} \gg m_{\nu_\e}$ to explain
the atmospheric and the solar neutrino observations naturally, and we consider 
only the tau and muon neutrino masses here.\footnote{
We keep three generation indices in our numerical 
calculation though we neglect indices for the first generation here.
}
Also, we assume that the Yukawa 
coupling and the Majorana masses for the right-handed neutrinos are real 
for simplicity. We parameterize two unitary matrices as 

\begin{eqnarray}
U_D= 
\left(
\begin{array}{cc} 
\cos\theta_D& \sin\theta_D \\
-\sin\theta_D& \cos\theta_D
\end{array}
\right),&&
U_M= 
\left(
\begin{array}{cc} 
\cos\theta_M&  \sin\theta_M \\
-\sin\theta_M& \cos\theta_M
\end{array}
\right).
\end{eqnarray}
The angle between $\nu_\mu$ and $\nu_\tau$ 
is related to $\theta_D$ and $\theta_M$ by
\begin{equation}
\theta_{\nu_\mu \nu_\tau } = \theta_D + \theta_M.
\end{equation}
However, in order to derive large $\theta_M$
we need  to fine-tune the 
independent Yukawa coupling constants and the 
mass parameters as explained below.
The neutrino mass matrix $(m_{\nu_L})$ in the second and  third generations
is written explicitly by
\begin{eqnarray}
(m_{\nu_L}) &=& 
\frac{1}{1-\frac{M_{\mu\tau}^2}{M_{\mu\mu}M_{\tau\tau}}}
\left(
\begin{array}{cc}
\frac{m_{\nu_\mu D}^2}{M_{\mu\mu}}&
-\frac{m_{\nu_\mu D}m_{\nu_\tau D}}{M_{\mu\tau}} \frac{M_{\mu\tau}^2}{M_{\mu\mu}M_{\tau\tau}}\\
-\frac{m_{\nu_\mu D}m_{\nu_\tau D}}{M_{\mu\tau}} \frac{M_{\mu\tau}^2}{M_{\mu\mu}M_{\tau\tau}}&
\frac{m_{\nu_\tau D}^2}{M_{\tau\tau}} 
\end{array}
\right).
\end{eqnarray}
Then, if the following relations are imposed, the neutrino mass
hierarchy  $m_{\nu_\tau} \gg m_{\nu_\mu}$ and $\theta_M\simeq \pi/4$ 
can be derived,
\begin{equation}
\frac{m_{\nu_\tau D}^2}{M_{\tau\tau}} 
\simeq \frac{m_{\nu_\mu D}^2}{M_{\mu\mu}}
\simeq \frac{m_{\nu_\mu D} m_{\nu_\tau D}}{M_{\mu\tau}}.
\label{relation}
\end{equation}
We need some mechanism to explain the relation among
the independent coupling constants and masses.  Also, 
if $m_{\nu_\tau D} \gg m_{\nu_\mu D}$, similar to the quark sector, the 
mixing angle $\theta_M$ is suppressed as
\begin{equation}
\tan 2 \theta_M \simeq
-2 \left(\frac{m_{\nu_\mu D}}{m_{\nu_\tau D}} \right)
  \left(\frac{M_{\mu\tau}}{M_{\mu\mu}}     \right),
\end{equation}
as far as the Majorana masses for right-handed neutrinos do not 
have stringent hierarchical structure as Eq.~(\ref{relation}).
Therefore, in the following discussion we assume that the 
large mixing angle between $\nu_\tau$ and $\nu_\mu$  comes from $\theta_D$. 

In the minimal supergravity scenario
the LFV masses for the left-handed sleptons may be induced radiatively
if the MSSM is extended to have the right-handed neutrinos \cite{BM}.
The SUSY breaking part of the lepton sector in the MSSMR is given as
\begin{eqnarray}
-{\cal{L}}_{soft}&=&
(m_{\tilde L}^2)_i^j {\tilde l}_{L}^{\dagger i}{\tilde l}_{Lj}
+(m_{\tilde e}^2)^i_j {\tilde e}_{Ri}^* {\tilde e}_{R}^j
+(m_{\tilde \nu}^2)^i_j {\tilde \nu}_{Ri}^* {\tilde \nu}_{R}^j
\nonumber \\
& &+ ( 
 A_l^{ij}   h_1 {\tilde e}_{Ri}^*  {\tilde l}_{Lj}
+A_\nu^{ij} h_2 {\tilde \nu}_{Ri}^*{\tilde l}_{Lj}
+ \frac12 b_{ij} {\tilde \nu}_{Ri}^* {\tilde \nu}_{Rj}^*
+ h.c.),
\label{m0}
\end{eqnarray}
where terms in the first line are the SUSY breaking masses for 
the left-handed and the right-handed sleptons, and those in the
second lines are 
the SUSY breaking terms associated with the supersymmetric Yukawa couplings
and masses. 
In the minimal supergravity scenario the SUSY breaking parameters 
at the gravitational scale ($M_{\rm G}$) are
\begin{equation}
(m_{\tilde L}^2)_i^j 
=(m_{\tilde e}^2)^i_j 
=(m_{\tilde \nu}^2)^i_j
= m_0^2 \delta^i_j,
\end{equation}
\begin{equation}
A_{\nu}^{ij}=f_{\nu}^{ij} a_0\ m_0 ,\, A_{l}^{ij}=f_{l}^{ij} a_0\ m_0,
\end{equation}
\begin{equation}
b_{ij}= M_{ij} b_0.
\end{equation}

These relations for the SUSY breaking parameters are unstable for the
radiative  
correction, and the LFV masses for the left-handed sleptons are generated
by the LFV Yukawa interaction for the neutrinos. Here, we assume that 
the Yukawa coupling constants for right-handed 
neutrinos in Eq.~(\ref{yukawacoupling}) have hierarchical structure 
$(f_{\nu_\tau}\gg f_{\nu_\mu} \gg f_{\nu_\e}$). In this case, the Yukawa 
interaction reduces only $(m_{\tilde L}^2)_\tau^\tau$  at the 
right-handed neutrino scale ($M_{\nu_R}$) as
\begin{eqnarray}
(m_{\tilde L}^2) &=&
\left(
\begin{array}{cc}
\bar{m}^2&\\
&\bar{m}^2-\Delta m^2
\end{array}
\right).
\end{eqnarray}
We consider only the left-handed slepton masses in the 
second and third generations here, again.
Here $\bar{m}^2$ is evaluated by one-loop level renormalization group (RG)
equations as 
\begin{eqnarray}
\bar{m}^2 &=& m_0^2 
+ \frac32 M_2^2 (M_{\nu_R})
\left(
\left(
\frac{g^2_2(M_G)}{g^2_2(M_{\nu_R})}
\right)^2
-1
\right)
+ \frac1{22} M_1^2 (M_{\nu_R})
\left(
\left(
\frac{g^2_Y(M_G)}{g^2_Y(M_{\nu_R})}
\right)^2
-1
\right)
\nonumber\\
&&
\label{m2}
\end{eqnarray}
where $M_2$ and $M_1$ are the wino and bino masses.
Here $\bar{m}^2$ becomes larger than $m_0^2$ due to the gauge interactions.
Assuming the GUT relation for the gaugino masses,  
$\bar{m}^2-m_0^2$ is almost as large as square of the wino mass at the weak
scale. Though we need numerical calculation in order to evaluate
$\Delta m^2$ by the RG equations, it becomes 
\begin{eqnarray}
\Delta m^2  &=& 
\frac{1}{4\pi^2} 
f_{\nu_\tau}^2 (3 +a_0^2) m_0^2
\log\frac{M_{\rm G}}{M_{\nu_R}},
\label{eq:deltam}
\end{eqnarray}
at the logarithmic approximation of one-loop level. 
In Eq.~(\ref{eq:deltam}) we ignore the radiative 
correction by the Yukawa couplings for charged leptons. 
The mass-square difference $\Delta m^2$ is proportional to $m_0^2$.

Below the right-handed neutrino scale, 
it is convenient to take basis of lepton where the Yukawa coupling constants 
for charged leptons are diagonal,
\begin{eqnarray}
W_{\rm MSSM}&=& 
 f_{l_i} H_1 E^c_i L_i
+ \frac{2 m_{\nu_k}}{v^2 \sin^2\beta} (U_M U_D)_{ki}(U_M U_D)_{kj} L_i L_j H_2 H_2,
\label{yukawacoupling1}
\end{eqnarray}
so that the supersymmetric Yukawa couplings are lepton-flavor conserving.
In this basis, $(m_{\tilde L}^2)$ becomes 
\begin{eqnarray}
(m_{\tilde L}^2)  &=& 
U_D^\dagger 
\left(
\begin{array}{ccc}
\bar{m}^2&\\
&\bar{m}^2-\Delta m^2
\end{array}
\right)
U_D.
\end{eqnarray}
The LFV off-diagonal terms in $(m_{\tilde L}^2)$ are 
controlled by $U_D$ and $\Delta m^2/m^2$.
It is important that if the $m_0$ is comparable to or smaller than $M_2$ and 
$M_1$, the LFV in the left-handed slepton mass matrix is suppressed. 

Similarly, 
the radiative correction generates the off-diagonal components of 
$A_l$. Since $A_l$ is proportional to the charged lepton Yukawa coupling
constants, it is  not relevant to our following discussion.

We parameterize sneutrino mass matrix by 
$\sin 2 \theta_{\tilde{\nu}}$ 
and mass splitting between two non-electron sneutrinos 
$\Delta m_{\tilde{\nu}}$ as
\begin{eqnarray}
(m_{\tilde{\nu}})= 
\left(
\begin{array}{cc} 
\cos\theta_{\tilde{\nu}}&-\sin\theta_{\tilde{\nu}}\\
\sin\theta_{\tilde{\nu}}& \cos\theta_{\tilde{\nu}}
\end{array}
\right)
\left(
\begin{array}{cc} 
\bar{m}_{\tilde{\nu}}^2& 0\\
0& (\bar{m}_{\tilde{\nu}} - \Delta m_{\tilde{\nu}})^2
\end{array}
\right)
\left(
\begin{array}{cc} 
\cos\theta_{\tilde{\nu}}& \sin\theta_{\tilde{\nu}} \\
-\sin\theta_{\tilde{\nu}}& \cos\theta_{\tilde{\nu}}
\end{array}
\right).
\end{eqnarray}
They are shown as 
functions of the right-handed neutrino mass scale in Fig.~(1). This is
derived by  
solving the RG equations at one-loop level numerically.\footnote{
Detailed formulae are given in Ref.~\cite{HMTY}.}        
Here, we take $m^2_{\nu_\tau}=0.005\ \mbox{eV}^2$, $\theta_D=\pi/4$, 
$\bar{m}_{\tilde{\nu}}=180$GeV, and $\tan\beta=3$, 10, 30. For simplicity, 
the gaugino masses in the MSSM
are given by the GUT relation, and the 
lightest chargino mass is fixed to 100GeV. 
If $f_{\tau}$ is negligible, $\Delta m_{\tilde{\nu}}$  depends on 
$f_{\nu_\tau}^2$ (or the right-handed neutrino scale, 
$M_{\nu_R}={f_{\nu_\tau}^2 v^2 \sin^2\beta}/{2m_{\nu_\tau}}$)
and $\theta_{\tilde{\nu}}$ is almost as large as $\theta_D$.
On the other hand, when $f_{\tau}$ is larger than  $f_{\nu_\tau}$, 
$\theta_{\tilde{\nu}}$ becomes smaller since $\Delta m_{\tilde{\nu}}$ is determined 
by  $f_{\tau}$, not $f_{\nu_\tau}$. Notice that the magnitude of the LFV
processes depends on $\sin 2 \theta_{\tilde{\nu}}\Delta m_{\tilde{\nu}}$.
In both of the region $\sin 2 \theta_{\tilde{\nu}}\Delta m_{\tilde{\nu}}$
is almost proportional to $f_{\nu_\tau}^2$ (or $M_{\nu_R}$), and it is not 
so sensitive to $f_{\tau}$ (or $\tan\beta$). 
For $M_{\nu_R}\gsim 10^{12}$GeV (which corresponds to 
$f_{\nu_\tau} \gsim f_{\tau}$ and  $\tan\beta \ge 3$) we can see 
$\sin2\theta_{\tilde{\nu}} \Delta m_{\tilde{\nu}} \gsim 0.1$GeV in the plot. 
This region may be accessible in future lepton colliders as we will show later. 

In Fig.~(2) the branching ratio of 
$\tau\rightarrow\mu\gamma$ in this model is shown. The parameters in 
this figure are the same as in Fig.~(1). 
At large $\tan\beta$ the branching ratio of
$\tau\rightarrow\mu\gamma$ is enhanced, and the line for $\tan\beta=30$ can 
reach to the present experimental bound \cite{PDG}.
This is because the matrix element of  $\tau\rightarrow \mu\gamma$ is almost 
proportional to $\tan\beta$ due to 
the dipole structure \cite{HMTY}. At the $B$ factories in KEK or 
SLAC the branching ratio of $\tau\rightarrow\mu\gamma$  may be probed at 
the level  of  $10^{-(7-8)}$. However, they are not sensitive enough to probe 
the LFV masses for sleptons  in low $\tan\beta$ or small $f_{\nu_\tau}$.
It is important to search for the LFV by direct production of the left-handed 
slepton in future colliders.

\section{Cross Sections for the LFV processes in lepton collisions}

In this section we present the formula for the cross sections of the 
LFV processes induced by the productions of the left-handed sleptons
in the second and third generations, and show the expected cross sections
of the LFV events. 

The characteristic LFV  signals depend on SUSY particle 
mass spectrum. 
In this article we take the SUSY particle mass 
spectrum expected in the minimal supergravity scenario. In  the scenario it 
is expected that 
the LSP is the bino-like neutralino ($\tilde{\chi}_1^0$), and that the second 
lightest neutralino ($\tilde{\chi}_2^0$) and the lightest chargino 
($\tilde{\chi}_1^-$) are wino-like. The LFV signals depend on whether
the left-handed sleptons are heavier or lighter than the wino-like chargino
and neutralino.

\begin{itemize}
\item 
{\it Case}.~(I): $m_{\tilde{l}_L}$, $m_{\tilde{\nu}} > m_{\tilde{\chi}_1^-}$, 
$m_{\tilde{\chi}_2^0}$

In this case the left-handed sleptons are heavy
enough to decay into two-bodies, the wino-like chargino or neutralino 
and lepton.
The signals are $\tau\mu+ 4jets + \E$, 
$\tau\mu l + 2jets+ \E$, and 
$\tau\mu ll + 2jets + \E$ $(l=\e,\mu,\tau)$,
where the missing energy $\E$ comes from the LSP's.
The main standard model BG comes from the 
$Z^0W^+W^-$ production process, and the BG cross section is small. The 
significance of the signal for the BG's may be large if the sleptons 
have the sizable LFV masses. 

\item 
{\it Case}.~(II): $m_{\tilde{l}_L}$, $m_{\tilde{\nu}} < m_{\tilde{\chi}_1^-}$, 
$m_{\tilde{\chi}_2^0}$

The left-handed sleptons  decay dominantly  into the LSP.
The signal is 
$\tau\mu +\E$, which comes from the charged slepton pair production. 
The main SM background comes from the $W$ boson pair 
production, which reaches to at least several fb in LC's even after the 
 cuts. 
Then, it is hard to get  significant signal over the background in LC's.
We will need efforts on reducing the background  in order to study this case 
in LC's. Also, in our model the LFV tends to be suppressed since the gaugino 
masses at the gravitational scale are larger than $m_0$ to realize 
this mass spectrum. 
Muon colliders may give a chance to study the LFV even 
in this case since the left-handed smuon pair production cross section may 
reach to 100fb due to $t$-channel neutralino exchange.  
In this article we do not study this case further since 
in the MSSM and the minimal supergravity scenario there is no region for 
$m_{\tilde{l}_L}$, $m_{\tilde{\nu}}<m_{\tilde{\chi}_1^-},$ 
$m_{\tilde{\chi}_2^0}<250$GeV and $\tan\beta\lsim$10, allowed by the 
experimental constraints.\footnote{
We include a constraint from $b\rightarrow s \gamma$ which excludes
the large region for $\mu<0$ \cite{GO}. 
}
 
\end{itemize}

Let us present formula for the cross 
section of $\tau^+ \mu^-  + \tilde{\chi}_A^-\tilde{\chi}_B^+$ $(A,B=1,2)$ 
through the sneutrino pair production in lepton collisions with the 
center-of-mass energy $\sqrt{s}$.  The subsequent decays of charginos 
lead to the signals  $\tau^+\mu^- + 4jets + \E$ and so on.
The amplitude of this process is given 
as \cite{ACFH2}
\begin{eqnarray}
{\cal M} &=& \sum_{X,Y} {\cal M}_{XY}(k^2,\bar{k}^2)
\nonumber\\
&&
\times \frac{i}
{k^2 - m_{\tilde{\nu}_X}^2 + i m_{\tilde{\nu}_X}\Gamma_{\tilde{\nu}_X}}
{\cal M}_{XB\mu}(k^2)
\times \frac{i}
{\bar{k}^2 -m_{\tilde{\nu}_Y}^2 + i m_{\tilde{\nu}_Y}\Gamma_{\tilde{\nu}_Y}}
{\cal M}_{YA\tau}^\dagger (\bar{k}^2),
\end{eqnarray}
where $X$ and $Y$ are indices for the mass eigenstates of sneutrino.
The momenta $k$ and $\bar{k}$ are for the sneutrino and the anti-sneutrino. 
Here, ${\cal M}_{XY}$ and ${\cal M}_{XAi}$
are amplitudes for production of the $X$-th and the $Y$-th off-shell 
sneutrinos and 
that for the $X$-th off-shell sneutrino decay into $A$-th chargino and $i$-th 
lepton. The explicit formulae are given in Appendix C.\footnote{
Here, we omit the indices for the helicities of the initial beams while 
we show them explicitly in Appendix C.
}
From this equation, we get the cross section as
\begin{eqnarray}
\sigma &=& 
\int  A_{XX^\prime}(k^2) dk^2
\int  A_{YY^\prime}(\bar{k}^2) d\bar{k}^2
\nonumber\\
&&
\times\sum_{X Y X^\prime Y^\prime}
\sigma_{XYX^\prime Y^\prime} (k^2,\bar{k}^2)
\times  {\rm Br}_{XX^\prime B\mu} (k^2) 
\times  {\rm Br}_{YY^\prime A\tau}(\bar{k}^2).
\end{eqnarray}
Here the off-shell sneutrino pair production cross section 
($\sigma_{XYX^\prime Y^\prime} (k^2,\bar{k}^2)$)  and the branching
ratio for off-shell sneutrino decay to the $A$-th chargino 
(${\rm Br}_{XYAi} (k^2)$) are 
\begin{eqnarray}
\sigma_{XYX^\prime Y^\prime} (k^2,\bar{k}^2)
\equiv
\int\frac1{2s} 
{\cal M}_{XY}(k^2,\bar{k}^2)
{\cal M}_{X^\prime Y^\prime}^{\dagger}(k^2,\bar{k}^2)
d\Omega_2,
\\
{\rm Br}_{XYAi} (k^2)
\equiv
\int\frac1{2[m\Gamma]_{\tilde{\nu}_X\tilde{\nu}_Y}} 
     {\cal M}_{XAi}(k^2)
     {\cal M}_{YAi}^{\dagger} (k^2)
d\Omega_2,
\end{eqnarray}
where
$[m\Gamma]_{XY}
= (m_{\tilde{\nu}_X}\Gamma_{\tilde{\nu}_X} 
 + m_{\tilde{\nu}_Y}\Gamma_{\tilde{\nu}_Y})/2$,
and $d\Omega_2$ is the two-body phase space integral. The 
function of the momentum for the sneutrino $A_{XY}(k^2)$ is
\begin{equation}
A_{XY}(k^2) \equiv \frac{1}{2\pi}\frac{2 [m\Gamma]_{XY}}
{(k^2 - m_{\tilde{\nu}_X}^2 + i m_{\tilde{\nu}_X}\Gamma_{\tilde{\nu}_X})
(k^2 - m_{\tilde{\nu}_Y}^2 - i m_{\tilde{\nu}_Y}\Gamma_{\tilde{\nu}_Y})}.
\end{equation}
For numerical calculations, we take only singular parts in this function as
\begin{equation}
A_{XY}(k^2) = \frac{1}{1+i x^{(\tilde{\nu})}_{XY}}
\frac{\delta(k^2-m_{\tilde{\nu}_X}^2)+\delta(k^2-m_{\tilde{\nu}_Y}^2)}{2},
\end{equation}
where 
\begin{equation}
x^{(\tilde{\nu})}_{XY} \equiv \frac{m_{\tilde{\nu}_X}^2-m_{\tilde{\nu}_Y}^2}
                   {2 [m\Gamma]_{XY}}.
\end{equation}
This function gives  correct cross section  in the interference region 
$m_{\tilde{\nu}_X}^2-m_{\tilde{\nu}_Y}^2 \lsim 2 [m\Gamma]_{XY}$ and 
the decoherent region 
 $m_{\tilde{\nu}_X}^2-m_{\tilde{\nu}_Y}^2 \gg 2 [m\Gamma]_{XY}$.
In order to evaluate the cross section in the intermediate 
region, we have to calculate the full one-loop correction 
to this process. However, that is out of scope of our paper.

By repeating the above procedure, we can get the cross sections
for $\tau^+\mu^-  + \tilde{\chi}_A^0\tilde{\chi}_B^0$ $(A,B=1-4)$ 
through the charged slepton pair production. Finally, we need to multiply 
the branching ratios of charginos and neutralinos to the cross sections
in order to get the cross sections for the LFV signals, such as
$\tau^+\mu^- + 4jets + \E$. We present the formula
for the decay widths of charginos and neutralinos in Appendix D.

Now, we present the cross sections of the signals at  LC's and 
$\mu^+\mu^-$ colliders. First, we discuss them in LC's. 
 In Fig.~(3) we present the cross section of  
$\tau^+\mu^-  + \tilde{\chi}_1^- \tilde{\chi}_1^+$ induced by the 
sneutrino pair production and that of  $\tau^+\mu^- +  \tilde{\chi}_2^0 
\tilde{\chi}_2^0$ induced by the charged slepton pair production.
The horizontal line is the mass splitting between two non-electron sneutrinos
$(\Delta m_{\tilde{\nu}})$
and the vertical line is the mixing angle ($\sin 2 \theta_{\tilde{\nu}}$).
Here, we take  the center-of-mass 
energy 500GeV,
the heaviest sneutrino (charged slepton) mass 180(194)
GeV,  the 
lightest chargino mass 100GeV and $\tan\beta=3$. For simplicity, we assume only the MSSM here, 
and the other SUSY particle masses are determined by the minimal supergravity 
scenario with the GUT relation for the gaugino masses and the radiative 
breaking condition of SU(2)$_L\times$U(1)$_Y$  with the Higgsino mass 
$(\mu)$ positive. For this sample parameter set the wino-like neutralino 
and the left-handed smuon masses are given in Table~(\ref{tab:parameter}).
In Tables~(\ref{tab:lcross}) and (\ref{tab:ldecay}), we show the charged
slepton and the sneutrino pair production cross sections in the LC and
the branching ratios  in the limit of zero flavor mixing. 
In order to obtain the signal cross sections for $\tau^+\mu^- + 4 jets +
\E$ and so on, the branching ratios of the chargino and the neutralino should be 
known precisely. Since they depend on details of the SUSY breaking mass 
spectrum, they should be determined experimentally. 
We present the branching ratios of the wino-like chargino 
and neutralino in the above assumption in Table~(\ref{tab:br}).

The widths of sleptons  are about 1.2 GeV.
The signal cross sections are not strongly suppressed 
if the mass difference $\Delta m_{\tilde{\nu}} \gsim 1$ GeV
as can be seen from Fig.~(3).
Therefore it may be possible to search for the LFV experimentally even 
if the mass 
difference $\Delta m_{\tilde \nu} \simeq 1$ GeV.

In Fig.~(3), the cross sections of  $\tau^+\mu^- + \tilde{\chi}_1^- \tilde{\chi}_1^+$ 
and  $\tau^+\mu^- +  \tilde{\chi}_2^0 \tilde{\chi}_2^0$ are comparable. 
The reason can be understood from the charged slepton and sneutrino pair
production cross sections in $\e^+\e^-$ collisions 
and the branching ratios in the limit of zero flavor 
mixing in 
Tables~(\ref{tab:lcross}) and (\ref{tab:ldecay}). The production cross 
sections for $\tilde{\mu}_L$ and $\tilde{\tau}_L$ are larger than those for 
$\tilde{\nu}_\mu$ and $\tilde{\nu}_\tau$ since $\tilde{\mu}_L$ and 
$\tilde{\tau}_L$ have QED charges. On the other hand,  since $\tilde{\chi}_1^-$
and $\tilde{\chi}_2^0$ are wino-like, the left-handed sleptons 
decay mostly  to $\tilde{\chi}^-_1$ and $\tilde{\chi}^0_2$, and  
${Br}(\tilde{f}\rightarrow f'\tilde{\chi}^{\pm})$ is larger than 
${Br}(\tilde{f}\rightarrow f \tilde{\chi}^0_2)$ (see Table~(3)).\footnote{
In pure wino limit ${Br}(\tilde{f}\rightarrow f'
\tilde{\chi}^{\pm}):$ ${Br}(\tilde{f}\rightarrow f \tilde{\chi}^0_2)
\sim$ $2:1$. However, in Table (3) the left-handed charged slepton decay into
$\tilde{\chi}^0_1(\tilde{\chi}^0_2)$ is suppressed (enhanced) while
the sneutrino decay into $\tilde{\chi}^0_1(\tilde{\chi}^0_2)$ is enhanced 
(suppressed). This comes from interference between wino- and bino-components
in the neutralinos.
}
Then, the cross section for $\tau^+\tau^-(\mu^+\mu^-) 
+\tilde{\chi}_1^{-}\tilde{\chi}_1^{+}$ is comparable to that for 
$\tau^+\tau^-(\mu^+\mu^-) +\tilde{\chi}_2^{0}\tilde{\chi}_2^{0}$ in the 
limit of zero flavor mixing. 
When $\tilde{\mu}_L$ and $\tilde{\tau}_L$, $\tilde{\nu}_\mu$ and 
$\tilde{\nu}_\tau$ have sizable mass difference, the cross section for  
$\tau^+\mu^- + \tilde{\chi}_1^- \tilde{\chi}_1^+$  from the sneutrino 
production may reach to 2.8fb, and that for 
$\tau^+\mu^- + \tilde{\chi}_2^0 \tilde{\chi}_2^0$ goes to 2.3fb at 
$\tan\beta=3$.
Then, from Table~(\ref{tab:br}), the cross sections for the 
signals are given approximately as
\begin{eqnarray}
\sigma(\tau^+\mu^- + 4jets + \E)
&\simeq& 
2\chi_{\tilde{\nu}}
(3-4\chi_{\tilde{\nu}})
\sin^2 2\theta_{\tilde{\nu}} \times 1.2 \fb,
\nonumber\\ 
\sigma(\tau^+\mu^- l^{\pm}+ 2jets+ \E)
&\simeq&
2\chi_{\tilde{\nu}}
(3-4\chi_{\tilde{\nu}})
\sin^2 2\theta_{\tilde{\nu}} \times 0.22 \fb,
\nonumber\\
\sigma(\tau^+\mu^- l^+l^- + 2jets + \E)
&\simeq&
2\chi_{\tilde{\nu}}
(3-4\chi_{\tilde{\nu}})
\sin^2 2\theta_{\tilde{\nu}} \times 0.15 \fb,
\label{eecollider}
\end{eqnarray}
where $\chi_{\tilde{\nu}}$ is given as
\begin{equation}
\chi_{\tilde{\nu}}\equiv
\frac{(x^{(\tilde{\nu})}_{12})^2}
{2(1+(x^{(\tilde{\nu})}_{12})^2)}.
\end{equation}

Next, in Fig.~(4) we show the cross sections of  
$\tau^+\mu^- + \tilde{\chi}_1^- \tilde{\chi}_1^+$ 
and $\tau^+\mu^- + \tilde{\chi}_2^0 \tilde{\chi}_2^0$ 
when $\tan\beta=10$. The other parameters are the same as in Fig.~(3).
The cross section of $\tau^+\mu^-+  \tilde{\chi}_2^0 \tilde{\chi}_2^0$ 
induced by the charged slepton pair production is drastically changed
compared to that for $\tan\beta=3$, while 
that of $\tau^+\mu^- + \tilde{\chi}_1^- \tilde{\chi}_1^+$ induced by 
the sneutrino pair production is not so.

The nontrivial  dependence on $\Delta m_{\tilde{\nu}}$ can 
be explained by the left-right mixing term of the 
stau's. 
The charged slepton mass matrix among
$\tilde{\mu}_L$, $\tilde{\tau}_L$, and $\tilde{\tau}_R$
is given by
\begin{equation}
\left(
\begin{array}{ccc}
m_{L22}^2 & m_{L23}^2 & 0\\
m_{L32}^2 & m_{L33}^2 & m_{LR33}^2\\
0         & m_{LR33}^2& m_{R33}^2 
\end{array}
\right),
\end{equation}
where $m_{LR23}^2$ is ignored. Each component is given in Appendix B.
As explained in Section 2, $ m_{L22}^2 \gsim m_{L33}^2 \gsim m_{R33}^2$ 
is expected in the minimal supergravity scenario, as far as the strong Yukawa
coupling for neutrino does not reduce $m_{L33}^2$ below $m_{R33}^2$. After 
diagonalizing the submatrix for the left-handed and right-handed stau's, 
this mass matrix becomes 
\begin{equation}
\left(
\begin{array}{ccc}
m_{L22}^2 & m_{L23}^2 \cos\theta_{\tilde{\tau}} & -m_{L23}^2 \sin\theta_{\tilde{\tau}}\\
m_{L23}^2\cos\theta_{\tilde{\tau}}              & m_{\tilde{\tau}_2}^2 & 0\\
-m_{L23}^2 \sin\theta_{\tilde{\tau}}            & 0        & m_{\tilde{\tau}_1}^2 
\end{array}
\right),
\end{equation}
where $m_{\tilde{\tau}_1}$ and  $m_{\tilde{\tau}_2}$ are masses of the stau's
and $\theta_{\tilde{\tau}}$ is the mixing angle. They are given as 
\begin{eqnarray}
m_{\tilde{\tau}_{1/2}}^2 
&=& 
\frac12 \left( m_{L33}^2 +m_{R33}^2 
\mp 
\sqrt{(m_{L33}^2 -m_{R33}^2)^2+ 4  (m_{LR33}^2)^2}
\right),
\nonumber\\
\tan2\theta_{\tilde{\tau}} &=& \frac{2m_{LR33}^2}{m_{L33}^2 -m_{R33}^2}.
\end{eqnarray}
When $(m_{L33}^2 -m_{R33}^2)$, which is almost as large as square of 
the wino mass in the minimal supergravity scenario, is larger than 
$|m_{LR33}^2|$,
the stau masses are given as
\begin{eqnarray}
m_{\tilde{\tau}_{1/2}}^2 &=& 
m_{R33/L33}^2 \mp \frac{(m_{LR33}^2)^2}{m_{L33}^2 -m_{R33}^2}.
\end{eqnarray}
If the left-right mixing for stau's $(m_{LR33}^2)$ is negligible, 
$\tilde{\tau}_2$ is lighter than $\tilde{\mu}_L$ with finite 
$\Delta m_{\tilde{\nu}}$. However, when the Higgsino mass or $\tan\beta$ 
is larger,  
$|(m_{LR33}^2)|$ is non-negligible, and $\tilde{\tau}_2$ is larger. If 
$\tilde{\tau}_2$ becomes more 
degenerate in the masses with $\tilde{\mu}_L$, the mixing angle is 
enhanced. The cross section is maximum when
$m_{L22}^2 -m_{L33}^2 \simeq (m_{LR33}^2)^2/(m_{L33}^2 -m_{R33}^2)$,
which corresponds to $\Delta m_{\tilde{\nu}} \simeq 4$ GeV
in Fig.~(4).
If the Higgsino mass or $\tan\beta$  is much larger and  $|m_{L33}^2 -m_{R33}^2|$ 
is smaller than $|m_{LR33}^2|$, the stau masses become 
\begin{eqnarray}
m_{\tilde{\tau}_{1/2}}^2 &=& 
\frac12 (m_{L33}^2+m_{R33}^2) \mp |m_{LR33}^2|.
\end{eqnarray}
In this case degeneracy between masses of $\tilde{\tau}_2$ and 
$\tilde{\mu}_L$ is more accidental, and usually the mixing is reduced 
effectively \cite{HT}. 
This kind of  behavior does not exist in the sneutrino production since 
the right-handed partners for neutrino  decouple at low energy.  We 
can extract directly the information on  the LFV in the SUSY breaking mass
terms for the left-handed sleptons by studying
the signal  $\tau\mu l + 2jet+ \E$ 
induced by sneutrino production.

Finally, we discuss the cross section for the signal at $\mu^+\mu^-$ colliders.
The production cross sections for left-handed sleptons 
in the second generation are much larger than the others due to $t$-channel 
exchanges of charginos or neutralinos. In Table~(\ref{tab:lcrossm}) we
show the production cross sections  
for slepton  at $\mu^+\mu^-$ colliders  in the
limit of zero  flavor mixing. Here, we take the sample parameter set for
$\tan\beta=3$  
in Table~(\ref{tab:parameter}).  
The $\tilde{\nu}_\mu$  pair production cross section reaches to
almost 1pb. On the other hand, that for $\tilde{\mu}_L$ is at most 150fb.
The destructive interference between $t$- and $s$-channels 
reduces  production of $\tilde{\mu}_L$ while the constructive 
interference enhances  the production cross section for  $\tilde{\nu}_\mu$.

In Fig.~(5) we show the cross sections for
$\tau^+\mu^- + \tilde{\chi}_1^-\tilde{\chi}_1^+$ by sneutrino production 
and $\tau^+\mu^- + \tilde{\chi}_2^0\tilde{\chi}_2^0$  by the charged
slepton production 
at $\mu^+\mu^-$ colliders. The parameter set is the same as in Fig.~(3). 
The cross section for $\tau^+\mu^- +  \tilde{\chi}_1^- \tilde{\chi}_1^+$ 
by the sneutrino pair production reaches to about 75fb with full mixing, 
while that for 
$\tau^+\mu^- + \tilde{\chi}_2^0 \tilde{\chi}_2^0$ by the charged
slepton slepton is at most 5.6fb due to the destructive interference. As
a result, the  cross 
sections for the signals are approximately
\begin{eqnarray}
\sigma(\tau^+\mu^- + 4jets + \E)
&\simeq& 
4 \chi_{\tilde{\nu}}
  \sin^2 2\theta_{\tilde{\nu}}
 \times 30 \fb
\nonumber\\ 
\sigma(\tau^+\mu^- l^{\pm} + 2jets+ \E)
&\simeq&
4 \chi_{\tilde{\nu}}
\sin^2 2\theta_{\tilde{\nu}}
 \times 6 \fb,
\nonumber\\
\sigma(\tau^+\mu^- l^+l^- + 2jets + \E)
&\simeq&
4 \chi_{\tilde{\nu}}
\sin^2 2\theta_{\tilde{\nu}} 
\times  0.4\fb.
\end{eqnarray}
Here, we assume the $t$-channel dominance in the production processes
and we take a small LFV limit.
Compared to the cross sections at LC
 Eq.~(\ref{eecollider}), the
signal cross sections at $\mu^+\mu^-$ collider is larger by one order
of magnitude. A $\mu^+\mu^-$ collider is a powerful 
tool to study the LFV in the 
second  generation.

\section{Detecting the LFV at future $\e^+\e^-$ and $\mu^+\mu^-$ colliders }
\subsection{Overview of signals and backgrounds}
In this section, we discuss the detection of the LFV
due to the  $\tilde{\tau}_L-\tilde{\mu}_L$ or 
the $\tilde{\nu}_{\tau}-\tilde{\nu}_{\mu}$ mixing in 
$\tau\mu + 4 jets$ and 
$\tau\mu l+ 2jet$ modes with $\E$. The 
branching fractions to the two modes are large, therefore 
they are good for the LFV study at  future lepton colliders. The former 
signal is simple involving only two leptons in the final 
state, however, the mode receives the contribution 
from both the left-handed charged slepton and the sneutrino productions. 
The mixing structure of the charged sleptons is complicated involving 
the  left-right mixing of stau's; In the previous 
section we saw $\tan\beta$ dependence of the signal. 
The latter mode receives the contribution only 
from $\tnu$ production, while 
the signal tends to be diluted due to the third lepton $l$. 
The signal $\tau\mu ll + 2 jets + \sla{E}$ is clean,
though we do not discuss it in detail here since
the cross section is small.

The tau lepton pair production in the SM processes 
generally provides background for $\tilde{\tau}$-$\tilde{\mu}$ 
and $\tilde{\nu}_{\tau}$-$\tilde{\nu}_{\mu}$ mixings. 
For example, 
$Z^0W^+W^-$ production can be a background when $Z$ decays into 
$\tau^+\tau^-$ and one of the two tau leptons decays into $\mu$. 
The production cross section of $Z^0W^+W^-$ is about 15 fb 
if we require central production of gauge bosons;
$\vert \cos\theta_V\vert <0.8$ \cite{miya}. Therefore, 
$\sigma(l^+l^-\rightarrow $ 
$ Z^0(\rightarrow\tau\tau\rightarrow\mu\tau) W^+W^-)= 0.17$fb.  
For our sample parameter set for $\tan\beta=3$, $\tilde{\chi}_1^-$ and 
$\tilde{\chi}_2^0$
do not decay into an on-shell 
$W$ or $Z$ boson, therefore we ignore this background completely, assuming 
jet invariant mass cuts.\footnote{
Even if 
$\tilde{\chi}^-_1(\tilde{\chi}^0_2)
\rightarrow W^- (Z^0) \tilde{\chi}^0_1$ is open, 
we can remove the background  by requiring  4jet kinematics 
are  consistent to $Z^0W^+W^-$ production. 
$hWW$ production with $h\rightarrow \tau^+\tau^-$ or other 
Higgs boson production also must be taken into account, because 
the branching ratio into $\tau$ may not be negligible.
} Another background 
is $\tau^+\tau^- W^+W^-$ production where the tau lepton pair comes from 
initial state radiation or bremsstrahlung. The tau leptons
typically have very low transverse momenta,  and they can be 
rejected by requiring large $P_T$ \cite{NFT} to the $\tau$ 
candidate leptons or jets. 

Rejecting backgrounds from the SUSY particle productions is also important 
for  the LFV study. For $\e^+\e^-$ colliders, the decay 
of sneutrino (left-handed charged slepton) pairs into $\tau^-\tilde{\chi}_1^+$
($\tau^-\tilde{\chi}^0_2$), 
would be important backgrounds. Decay of one of two $\tau$'s
into $\mu$ produces $\mu\tau X$ events; 17\% of decaying $\tau$'s into $\mu$.
If we cannot reject the muon, and 
tau leptons are identified by only the hadronic decay 
$(Br(\tau\rightarrow hadrons)= 64\%)$, 
detection of the LFV at 
$\e^+\e^-$ colliders  may not be very promising. 
When 500 $\tau\tau X$ events are produced,
$500 \times 0.64\times 0.17 \times 2 \sim 110$  
events of them will be  the background  $\tau\mu X$ candidates
for 100 \% acceptance. 
Then  $3\sigma$ discovery of 
$\tilde{l}^+\tilde{l}^-$ or $\tilde{\nu}\tilde{\nu}^c\rightarrow$ 
$\tau\mu X$ requires more than 
$\sqrt{110}/Br(\tau\rightarrow hadron)\times 3\sim 50$ 
signal productions; $S/N$ is 0.52.
We need to find cuts to 
eliminate the background from $\tau\rightarrow \mu$ and/or to increase $\tau$ 
acceptance to study the LFV
phenomena at $\e^+\e^-$ colliders. 
In Section 4.2 we discuss cuts on the muon energy  $E_{\mu}$ and 
the impact parameter $\sigma_{IP}$ (distance between the $\mu$ track to
the interaction point), and  
estimate the probability to 
misidentify $\mu$ from $\tl\rightarrow\tau$ as $\tl\rightarrow\mu$ 
under the cuts.
The sensitivity for $\sin 2\theta_{\tilde{\nu}}$ and $\Delta m_{\tilde{\nu}}$
will be discussed in Section 4.3 for the $\tau\mu+ 4jets$ mode. 

For $\mu^+\mu^-$ colliders, study of the LFV  would  
be easier because of the  $\tnu_{\mu}\tnu_{\mu}^c$ 
and $\tilde{\mu}^+\tilde{\mu}^-$ productions dominate
 over the $\tnu_{\tau}\tnu_{\tau}^c$ and
$\tilde{\tau}^+\tilde{\tau}^-$ productions in the limit 
of the small LFV. In Section 4.3, we will compare the experimental 
sensitivity for the
 LFV at $\mu^+\mu^-$  colliders to that at $\e^+\e^-$ colliders. 
Because large statistics is expected for $\mu^+\mu^-$ colliders, 
we discuss the detection of the LFV in $\tau\mu l +2jets$ mode. 
Some modes are masked by SUSY backgrounds such  
as $\mu^+\mu^-\rightarrow\tnu\tnu^c\rightarrow
\mu^+\nu_{\mu}\tilde{\chi}_2^0(\rightarrow\tau^+\tau^-)\tilde{\chi}_1^-(\rightarrow 2jets)$. 

\subsection{Energy and impact parameter cuts}
In this section we discuss cuts to 
eliminate $\tau\tau X $ background 
to $\tau\mu X$ signal. A cut on the  muon energy $E_\mu$
must be very useful, and a cut on the impact parameter $\sigma_{IP}$
is the other possibility if a fine vertex detector is available.  

The tau lepton decays into $e^-$, $\mu^-$, $\pi^-$, $\rho^-$, etc.,
and 
the lifetime is $c\tau_{\tau}= 86.93\mu m$. The decay distributions 
depend on the spin of the initial and the final particles. The detailed 
discussion on that can be found in \cite{tau} and references therein. 

The decay $\tau^-\rightarrow\mu^-\nu_{\tau}\bar{\nu}_{\mu}$ 
involves two neutrinos, and then $E_{\mu}$ is significantly softer
than that of the tau lepton. In Fig.~(6), 
we show the energy distribution of
$\mu$ from decaying $\tau$ when the tau lepton comes from sneutrino 
pair production $\e^+\e^-\rightarrow\tnu\tnu^c$ and decay 
$\tilde{\nu} \rightarrow \tau^-\tchi_1^+$
($\tilde{\nu}^c \rightarrow \tau^+\tchi_1^-$). 
The $\tau$ energy distribution is flat 
between the two end points $E_{\tau}^{\rm min}$ and $E_{\tau}^{\rm max}$ ,
\begin{equation}
E^{\rm max(min)}_{\tau}=E^{C.M.}_{\tau}
(1\pm\beta_{\tilde{\tau}})\gamma_{\tilde{\tau}},
\end{equation}
where 
\begin{eqnarray}
E^{C.M}_{\tau}&=& \frac{m^2_{\tnu}-m^2_{\tchi^-_1}}{2 m_{\tnu}}.
\end{eqnarray}
Here, $\beta_{\tilde{\nu}}^2=1-4 m_{\tnu}^2/s$ and 
$\gamma_{\tilde{\nu}}^2=1/(1-\beta_{\tilde{\nu}}^2)$, and 
we ignore $m_{\tau}$.
For our sample parameter set for $\tan\beta=3$, $E_{\tau}^{\rm max}=146$ GeV
and  $E_{\tau}^{\rm min}=26.5$ GeV. The parent $\tau$ distribution is
also shown in the plot. 

The muon energy distribution has mild  dependence on $\tau$
polarization $P_{\tau}$, 
and it gets harder as $P_{\tau}$ is decreased. The solid line shows 
the distribution for $P_{\tau}=-1$, while the dashed line shows 
the distribution for $P_{\tau}=1$. For the case we concern, 
 $P_{\tau}\sim - 1$; notice that the LFV
occurs due to the mixing of left-handed sleptons 
in our model, and $\tilde{\chi}^-_1$ is gaugino-like in the minimal supergravity 
scenario, therefore the tau lepton is left-handed \cite{nojiri, NFT}. 

The energy distribution for the signal $\mu$ is flat between 
$E_{\mu}^{\rm max}$ and $E_{\mu}^{\rm min}$ and the end points are the 
same as $E_{\tau}^{\rm min}$ and $E_{\tau}^{\rm max}$. 
Therefore we can reduce the background $\tau\rightarrow\mu$ 
by requiring $E_{\mu}>E_{\mu}^{\rm min}$ without costing signal events; 
52\% of $\mu$ from the $\tau$ decay remains as the candidate of 
primary $\tilde{\nu}\rightarrow\mu$ decay if 
we require $E_{\mu}>E_{\mu}^{\rm min}$. Though increasing $E_{\mu}^{\rm cut}$ 
reduces signal, $S/N$ would be improved considerably. For
$E^{\rm cut}_{\mu}= 50 (70)$ GeV, 25\%(1\%) of $\tau\rightarrow\mu$
remains  and the $S/N$ ratio is  improved by a factor of 3(90) to 
that without the cut. 

We now consider a cut on   $\sigma_{IP}$.
The measurement of the distance $\sigma_{IP}$ using a vertex 
detector might  be  useful to reduce background. The $\mu$ from the signal  
$\e^+\e^-\rightarrow\tilde{l}^-\tilde{l}^+$ or $\tilde{\nu}\tilde{\nu}^c$ 
$\rightarrow\tau\mu\tchi\tchi$ must have a track which  points back toward the interaction 
point. By requiring signal event to have $\sigma_{IP} < 
\sigma_{IP}^{\rm cut}$ one can eliminate some backgrounds.

When a relativistic $\tau$ lepton 
is produced at an interaction point, it flies for length $l$ 
and  then decays. The decay distribution may be 
obtained by boosting the decay distribution at the $\tau$ rest frame. 
Let us denote the muon momentum at the $\tau$ rest frame as 
$(E^0, p^0_{T1}, p^0_{T2}, p^0_{//})$, where $p^0_{//}$ is the component 
parallel to the boosted direction. 
When $p^0_{//}$ is large and positive, 
the daughter $\mu$ energy is large in the laboratory 
 frame, namely, by ignoring 
$m_{\mu}$, 
\begin{eqnarray}
 E_{\mu} = (\beta_{\tau} p^0_{//} +  E^0)\gamma_{\tau}.
\end{eqnarray}
The angle between  $\mu$ and $\tau$ momenta in the laboratory frame is 
\begin{eqnarray}
\tan\theta = \frac{\vert p^0_{T}\vert }
{(p^0_{//} + \beta_{\tau} E^0)\gamma_{\tau}}.
\end{eqnarray}
Apparently, when the $\tau$ energy is large, $\theta$ is small. 
However the average $\tau$ flight length increases proportional 
to $\gamma_{\tau}$, typically $l\sim \gamma_{\tau}\beta_{\tau}c\tau_{\tau}$. 
Therefore the measurable quantity $\sigma_{IP}$ is 
\begin{eqnarray}
\langle\sigma_{IP}\rangle&=& \langle l\sin\theta\rangle
\cr
&\sim & c \tau_{\tau}\frac{\beta_\tau\vert p^0_{T}\vert}
{(p^0_{//} + \beta_{\tau} E^0)}.
\label{sigma}\end{eqnarray}
Notice that $\gamma_{\tau}$ dependence is now disappeared. 
In the  $\beta_{\tau}\sim 1$ limit, $\sigma_{IP}$ distribution 
depends only  on the $\mu$ decay distribution in the rest frame.  
Unless $\beta_{\tau}\ll1$, $\sigma_{IP}$ tends to be 
smaller than $c\tau_{\tau}$, because $E^0>\vert p^0_{T}\vert$. 
The $E_{\mu}$ and $\sigma_{IP}$ distributions may be obtained by
numerically convoluting the  distributions for $\tau$ decaying into $\mu$
with the parent $\tau$ energy distribution.
The detailed formula is given in 
the Appendix E.

Fine vertex resolution 
might be achieved at future $\e^+\e^-$ LC experiments.
For the JLC1 detector, a CCD vertex 
detector is proposed, and the impact parameter resolution $\delta\sigma$ 
is estimated as 
$\delta\sigma = 11.4\oplus 28.8/p\beta/\sin^{3/2}
\theta(\mu m)$  \cite{JLC}.  The 
recent NLC detector
study shows that $\delta\sigma=2.4\oplus 13.7/p\beta/\sin^{3/2}\theta(\mu m)$
is possible \cite{NLC}. 
For $\mu^+\mu^-$ colliders, the proper configuration of the detector 
is still under preliminary consideration. Large backgrounds
of electrons, photons, pions, and muon's are expected 
near the interaction region. Because  the large 
signal cross section is expected for muon colliders, 
we can expect  good experimental sensitivity to the LFV
without an IP cut.

It is important to have $\delta\sigma_{IP} \ll c\tau_{\tau}$,
as $\sigma_{IP}$ is very likely less than $c\tau_{\tau}$.
Notice that  the error on  the  $\sigma_{IP}$ is estimated about  
2.5$\mu m$ at $E_{\mu}\sim E^{\rm min}_{\mu}$ in  the NLC detector.
If the interaction point is well-known,  $\sigma^{\rm cut}_{IP}\sim 10\mu m$ 
sounds reasonable choice. One question is if the interaction
 point is determined accurately enough. 
The particles contained in the jets are generally 
softer than the parton energy, therefore the $\sigma_{IP}$ 
resolution of  each track in the jets is less accurate. 
The resolution of interaction point will be determined by combining the 
each track measurement.   The estimation of the resolution requires 
detailed detector studies, which are out of scope of this paper.
Even if the interaction point cannot be measured 
precisely, the beam position must be  very accurately known for LC's. 
One might try to measure the distance between 
the $\mu$ track and the beam axis, although it 
 generally gives poor results than the one given here.

In Fig.~(7), we show the energy distribution of $\mu$ from 
$\tnu\rightarrow\tau\rightarrow\mu$ decay requiring 
$\sigma_{IP}<\sigma^{\rm cut}_{IP}= 10$, 30, 50, $90\mu m$.  
The background decreases as one require 
a tighter $\sigma_{IP}$ cut. Notice that
the background is not reduced too much when $E_{\mu}$ 
is close to  $ E^{\rm max}_{\mu}$. In such a case 
as $\vert p^0_T\vert <<\vert p^0_{//}\vert$, the typical $\sigma_{IP}$ 
is smaller as can be seen from Eq.~(\ref{sigma}).  In Fig.~(8) 
we show $E_{\mu}$ distribution for $\sigma_{IP}<\sigma_{IP}^{\rm cut}$ 
as a ratio to $E_{\mu}$ distribution without the $\sigma_{IP}$ cut.
We see the moderate reduction of the background in the 
wide range of $E_{\mu}$, which helps to improve the $S/N$ ratio 
without reducing signal events. 
If we require $\sigma_{IP}$  is less than 10$\mu m$, 
the background is reduced by a factor of 6 at $E_{\mu}\sim 
E^{\rm min}_{\mu}$.

The  $\sigma_{IP}^{\rm cut}$  cut
is attractive in the sense  to increase $S/N$ 
without costing signal events. 
In Fig.~(9) we plot $p_{\mu}$, the probability  that 
$\tnu$ production and its decay 
into $\tau$ is mis-identified as $\tnu\rightarrow\mu\tchi$.
We assume that  $\mu$ from direct  
$\tilde{f}\rightarrow\mu\tchi$ decay is selected when 
following conditions are satisfied;
\begin{itemize}
\item $E^{\rm min}_{\mu}<E_{\mu}<E^{\rm max}_{\mu}$,
\item $\sigma_{IP}<\sigma^{\rm cut}_{IP}$.
\end{itemize}
%

Isolated hard low multiplicity jets will be identified as 
$\tau$.  We assume all the  
hadronic decay of $\tau$ are identified as $\tau$ candidates. 
In addition, the leptons with $\sigma_{IP}>\sigma^{\rm cut}_{IP}$ 
may be regarded as $\tau$ candidates, too. 
We define the $\tau$ identification 
probability $p_{\tau}$ as
\begin{equation}
p_{\tau}=0.64+ 0.35\times p,
\end{equation}
where $p$ is the probability that lepton from 
decaying $\tau$ has $\sigma_{IP}>\sigma^{\rm cut}_{IP}$ 
and $E_{l}>5$ GeV. 
\footnote{We need to veto soft jets or leptons to reject 
soft particles from jets. The hadronic $\tau$ decay product 
carries a substantial fraction of $\tau$ energy when the mass 
of the decay products is of the order of $m_{\tau}$, therefore 
effect of the energy cut is small. } 
We show $p_{\tau}$ in Table~(6) with $p_{\mu}$.

Isolation cuts for the leptons and  other cuts 
must be applied to select signal events. They may be taken care 
as an overall acceptance common to $\tau\tau$, $\mu\mu$ and 
$\tau\mu$ events.  We discuss them in the next subsection.

\subsection{$\tau\mu+ 4jets$ mode}  

We first discuss $\tau\mu+ 4jets$ mode with $\E$. The mode has large 
branching ratio, and the interpretation 
of signal is simple.  In the limit of zero flavor mixing, 
$\mu\mu +4jets$ event will be produced around 120fb for 
muon colliders, and 2.5fb for $\e^+\e^-$ colliders. The signal  
$\tau\mu +4jets$ candidates will be selected by 
the $E_{\mu}^{\rm cut}$ and the IP cuts discussed in the previous 
subsection, and 
we assume backgrounds dominantly come from $\tau^+\tau^- +4jets+\E$ 
production.  The background can be subtracted by 
simultaneously measuring the cross section.

The significance of the 
signal ${\cal S}$ is a function of accepted 
numbers of $\tau\mu +4jets+\E$ and $\tau\tau +4jets+\E$ 
events, $O_{\tau\mu}$ and $O_{\tau\tau}$. 
When statistics of the signal and the background is large, 
we can assume Gaussian distribution and 
\beq
{\cal S}&=&O_{sig}/\delta O_{sig},
\label{ot1}
\eeq
where 
\beq
(\delta O_{sig})^2&=& O_{\tau\mu}+ 4 (p_{\mu}/p_{\tau})^2 O_{\tau\tau}.
\label{ot2}
\eeq
Here $p_{\mu}$ and $p_{\tau}$  are the $\mu$ misidentification rate 
and the $\tau$ identification rate discussed before, and 
$O_{sig}= O_{\tau\mu}-O_{bg}$, where $O_{bg}$ is estimated 
as $O_{bg}= (p_{\mu}/ p_{\tau}) O_{\tau\tau}$.
The $O_{\tau\tau}$ and $O_{sig}$ 
are related to the signal cross section by 
\beq
&O_{\tau\tau}= &A_{\tau\tau} p_{\tau} p_{\tau} N_{\tau\tau},
\label{ot3}
\cr
&O_{sig}= &A_{\tau\mu} p_{\tau}(N_{\tau\mu}+N_{\mu\tau}), 
\label{ot4}
\eeq
and 
\beq
N^{e(\mu)}_{l_i l_j}&=& {\cal L} \times\sigma(\e^+\e^-(\mu^+\mu^-)
\rightarrow \tilde{l}^-\tilde{l}^+~{\rm or}~
\tilde{\nu}\tilde{\nu}^c\rightarrow l^+_i l^-_j 
+4jets+\sla{E}),
\label{ot5}
\eeq
where ${\cal L}$ is the integrated luminosity and $A_{ij}$ is 
the acceptance of $\tau\tau(\tau\mu)+ 4jets$ modes.

Notice that we add signals from the sneutrino
and the charged slepton productions. This is because 
$m_{\tilde{\nu}}\sim m_{\tilde{l}_L}$
and $m_{\tilde{\chi}^0_2}\sim m_{\tilde{\chi}^-_1}$, therefore 
$\tau\mu +4jets$ events 
from the charged slepton and sneutrino productions have similar kinematical
signature. The muon misidentification rate $p_{\mu}$ depends 
on primary $\mu$ criteria  which is 
specified by the  $E_{\rm \mu}^{\rm cut} $ and $\sigma^{\rm cut}_{IP}$.
The energy distribution of $\mu$ from
$\tilde{\nu}/\tilde{\l}\rightarrow$
$\tau\rightarrow\mu$ depends on the charged slepton and the sneutrino
masses, and  the chargino and  neutralino masses, 
therefore $p_{\mu}$  is a function of those masses.
However, in our model, the energy distributions of $\mu$ from 
different decay chains are similar, so  
we choose common $p_{\mu}$ for charged slepton and sneutrino backgrounds 
to estimate the experimental sensitivity. For 
LC's we take $p_{\mu}=0.02$ and $p_{\tau}=0.88$ for 
$\sigma^{\rm cut}_{IP}=10\mu m$.
For muon colliders, we use the value without the IP cut,
$p_{\mu}=0.09$ and $p_{\tau}=0.64$. 

The slepton mass uncertainty  is removed by
studying the end point energy of 
electron or muon from $\tilde{\nu}_{\e(\mu)}$ decay. 
The production cross section of $\tilde{\nu}_{\e(\mu)}$ is large 
at $\e^+\e^-$ ($\mu^+\mu^-$) colliders, and  one can determine the 
masses of sneutrino and chargino very precisely by the
method in Ref.~\cite{BMT,TSUKA}. 
The mass resolution of 2\% for the $\tilde{\nu}$ is 
straightforward if the cross section is of the order 
of 1  pb.  The $\tilde{\e}_L$ and $\tilde{\mu}_L$ 
productions should also give us the information on the left-handed charged 
slepton masses. 

No Monte Carlo (MC) study involving $\tilde{\tau}_L$ and $\tilde{\nu}_\tau$
has been done.  Therefore  we seek for a 
reasonable 
assumption for $A_{ij}$. Those acceptances are again in principle 
 different between the signals from $\tilde{\nu}$ 
and $\tilde{l}$ productions. However, we assumed that
they are universal because of the kinematical similarity 
of the mode. The MC study of the process 
$\tilde{\tau}_R^+\tilde{\tau}^-_R\rightarrow$ 
$\tau^+\tau^- \tilde{\chi}^0_1\tilde{\chi}^0_1$  has been done in
Ref.~\cite{NFT}
without assuming a vertex detector.  
In the paper, the tau jets are identified by suitable multiplicity cuts, 
jet invariant mass cuts, and no substantial reduction 
of the signal events coming from the $\tau$ identification
was found. However, the isolation cut from other jet activities for 
$\tau$ and $\mu$ candidates has to be required for our case, 
and it might reduce accepted events. 
On the other hand, the SM backgrounds for the signal of the left handed 
slepton production are substantially 
smaller than that for  the ${\tilde \tau}_R^+{\tilde \tau}_R^-$ production
case because of the multiple jets in the final state, and we do not need 
strong kinematical cuts applied in Ref.~\cite{NFT}.
Indeed, for the $\tilde{\nu}$ study in Ref.~\cite{BMT}, the acceptance 
is found around 50\%, higher than those typical 
for $\tilde{\chi}^-_1$ or $\tl_R$ studies, 30\%. We therefore 
assume 30\% as the overall acceptance $A$ to the signal. 

Now we are ready to estimate experimental sensitivity for the LFV 
process based on Eq.~(\ref{ot1}).
In Fig.~(10) we plotted the contours of constant
significance corresponding to  $3\sigma$ discovery
in $\sin2\theta_{\tilde{\nu}}$ 
and $\Delta m_{\tilde{\nu}}$ plain. 
We take the sample parameter set for $\tan\beta=3$ in Table~(1), and 
parameterize the sneutrino mass matrix by 
$\sin 2 \theta_{\tilde{\nu}}$ and 
the mass splitting between two non-electron sneutrinos
$\Delta m_{\tilde{\nu}}$.
We assume 
the integrated luminosity ${\cal L}=50$fb$^{-1}$, proposed  
one year luminosity of  LC's.
For the $\e^+\e^-$ colliders, the experimental sensitivity is limited by 
the small statistics of the signal. 
The parameter region with $\sin 2 \theta_{\tilde{\nu}}>0.5$ and 
$\Delta m_{\tilde{\nu}}>0.4$ GeV 
may be explored.
However, the  background 
is strongly suppressed by the factor of $p_{\mu}=0.02$
(corresponding to $\sigma^{\rm cut}_{IP}= 10\mu m$.) 
Provided accidental backgrounds (such as $\tau$ lepton off from 
a charm jet)
can be controlled, the sensitivity may be extended up to the region 
where $S/N\sim1$.(That corresponds to the region 
$\sin 2\theta_{\tilde{\nu}}
> 0.2$ and $\Delta m_{\tilde{\nu}} >0.1$GeV in our sample parameter). 

For the case of $\mu^+\mu^-$ collisions, the  signal rate is a factor 10 larger
than $\e^+\e^-$ collisions as shown in Table~(2) and (5).
Notice that $S/N$  in the small 
mixing region is enhanced due to the enhanced 
$\mu^+\mu^- +4jets+\E$ rate over the background $\tau^+\tau^- +4jets+\E$ 
rate. The region where
$\sin(2\theta_{\tilde{\nu}})>0.1$  
or $\Delta m_{\tilde{\nu}}>0.1$  GeV can be explored for 
the integrated luminosity ${\cal L}=50$fb$^{-1}$.

\subsection{$\tau\mu l+ 2jets$ modes}  
The signal $\tau\mu+ 4 jets+\E$, which we considered in the previous 
section,  involves many jets in the final state.  
Since large fraction of the $\tau$ leptons is  detected by 
the hadronic decays, and the background from $c\rightarrow\tau$ off 
the jet axis could be large.   This means that  
we might need very tight isolation cuts, resulting poor 
acceptance. 
Therefore it is useful to search for the LFV in $\tau\mu l+ 2jets$ modes.
Notice that the process is induced by  only $\tnu$ production 
as shown in 
Eq.~(\ref{signals}). The detection of the LFV in these modes might provide 
clean information on the sneutrino mass matrix. 

The search of $\tau\mu l +2jets$ modes 
at muon colliders would be promising, because the 
production cross section for $\tnu_{\mu}\tnu_{\mu}^c$ is enhanced.  
The total cross section for  $(\mu^+\mu^-\l +2jets)$ 
is 140 fb for our sample parameter set for $\tan\beta=3$ 
when the mixing is small. 
The cross section for $\tau^\pm\mu^\mp l  +2jets$ 
will be suppressed by a factor of $\sin^2
2\theta_{\tilde{\nu}}$ to the large  
cross section for $\mu^+\mu^- l+2jets$ 
when the sneutrino mass difference is 
large enough. 

There are several sources of SUSY backgrounds to the 
signals. The sneutrino and the charged slepton decays 
into neutrino cause $3l$ signal. The dominant one is 
\begin{eqnarray}
\tnu\tnu^c
&\rightarrow& \mu^- \bar{\nu}+\tchi_1^+\tchi_2^{0} 
\nonumber\\
&&
\left\{\begin{array}{c}
\tchi_1^+\rightarrow  2jets+\tchi_1^0\\
\tchi_2^{0}\rightarrow  \tau^+\tau^- +\tchi_1^0
\end{array}\right.
\ ( 6.6{\rm fb}),
\label{B1}\\
\tl^-\tl^+
&\rightarrow& 
\mu^-\bar{\nu}+\tchi_2^0\tchi_1^{+} 
\nonumber\\
&&
\left\{\begin{array}{c}
\tchi_2^0\rightarrow \tau^+\tau^-+\tchi_1^0\\
\tchi_1^{+}\rightarrow 2jets +\tchi_1^0
\end{array}
\right.
\ (2.8 {\rm fb}).
\label{B3}
\end{eqnarray}
Those are the  backgrounds from the charged slepton or the sneutrino 
decay into $\mu$, which have large cross sections 
in the limit of the small LFV. (The production cross sections 
in zero LFV limit are listed at the end of 
Eqs.~(\ref{B1},\ref{B3}).)

Notice that the energy distribution of $\mu$ from the primary slepton 
decay in Eqs.~(\ref{B1},\ref{B3}) is similar  to that of the signal, therefore 
it is not possible to reduce the background by the 
 $E_{\mu}$ cuts. The situation is different 
to the previous $\tau\mu +4jets $ study. For the previous case 
the background can be reduced
because of the soft energy spectrum of $\mu$ from the $\tau$ decay. 

Because of the background, the $\tau^{\pm}\mu^{\mp}\tau^{\mp}$ modes
and $\tau^{\pm}\mu^{\mp}\mu^{\mp} (\e^{\mp})$ are not 
promising.\footnote{
The background $\tau$ energy distribution from 
$\tchi^0$ decay may be softer than signal. However, 
the cut to  $E_{\tau}$ would not be too efficient because the 
 energy distribution is smeared by $\tau$ decay.
}
In the limit of the small LFV, $S/N$ for $\tau^{\pm}\mu^{\mp}\tau^{\mp}$
would not 
go too much above $1/2(\sin 2\theta_{\tilde{\nu}})^2\times$
$\sigma(\mu^+\mu^-\tilde{\chi}_1^+(\rightarrow\tau^+)\tilde{\chi}_1^-)\vert_{\rm no-mixing}/$
$\sigma(\nu\mu^-\tilde{\chi}_2^0(\rightarrow\tau^+\tau^-)\tilde{\chi}_1^+)\vert_{\rm
no-mixing}\sim 1.2(\sin 2\theta_{\tilde{\nu}})^2$. 
For $\tau^{\pm}\mu^{\mp}\mu^{\mp} $ 
and  $\tau^{\pm}\mu^{\mp}\e^{\mp}$ 
modes, the background 
will be suppressed by 
$\tau$ branching ratio into $\mu$ or $\e$,  however, those modes are 
not attractive compared to the modes we  discuss below.

The modes $\tau^{\pm} \mu^{\mp} l^{\pm}+2jets$ ($l\equiv\tau,\mu,\e$) 
do not suffer large background contribution from  
$\mu^{\mp}\nu(\bar\nu)\tchi^0\tchi^{\pm}$ production.
Therefore we discuss the modes rather carefully.

\begin{itemize}
\item 
$\tau^+\tau^+\mu^-+2jets$   mode 

For $\tau^+\tau^+\mu^-$ mode,  the signal comes 
from  sneutrino pair decay into $\tau^+\tchi_1^-\mu^-\tchi_1^+$ 
followed by 
decay of $\tchi_1^+$ into $\tau^+\nu\tchi^0_1$ and that of $\tchi_1^-$ 
into $2jets + \tchi^0_1$. Those three 
leptons in the final state must be identified correctly. 
Backgrounds are expected from 
\begin{eqnarray}
(A)\ \ \ 
\mu^+\mu^-\rightarrow &\tilde{l}^+\tilde{l}^-\  {\rm or}\  \tilde{\nu}\tilde{\nu}^c&
\rightarrow \tau^+\nu+\tchi^0_2\tchi^-_1
\cr 
&&
\left\{\begin{array}{c}
\tilde{\chi}^0_2\rightarrow\tau^+\tau^-(\rightarrow\mu^-)+\tchi_1^0\\
\tchi^-_1\rightarrow 2jets +\tchi_1^0
\end{array}
\right.
\end{eqnarray}
and 
\begin{eqnarray}
 (B)\ \ \ 
\mu^+\mu^-\rightarrow & \tilde{\nu}\tilde{\nu}^c &
\rightarrow \tau^+\tau^-(\rightarrow\mu^-) +  \tchi_1^+ \tchi_1^- 
\cr 
&&
\left\{\begin{array}{c}
\tchi_1^+\rightarrow\tau^+\nu_\tau +\tchi_1^0\\
\tchi_1^-\rightarrow 2jets+\tchi_1^0
\end{array}\right.
\end{eqnarray}
with misidentification of  $\tau^-\rightarrow \mu^-$ as primary $\mu^-$.
The background cross sections are small in the limit 
of zero slepton mixing since they are induced by the 
$\tilde{\tau}_L$  or $\tilde{\nu}_\tau$ production.
If the mixing is non-zero, 
the $t$-channel chargino or neutralino exchange contribution to them
is suppressed by a factor proportional to 
$\sin^2 \theta_{\tilde{\nu}}$ for (A), and 
$\sin^4 \theta_{\tilde{\nu}}$ for (B).
For background (A), $\tau$ from $\tchi^0_2$ decay  must 
have soft energy spectrum, therefore $\mu$ from the $\tau$ decay, too. 
We assume the  background (A) will be negligible if a moderate 
$E_{\mu}$ cut is applied.  The rejection of background 
(B) is analogous to the $\tau\mu +4jets$ mode. 

\item $\tau^+\e^+\mu^-+2jets$ mode

The background from
\begin{eqnarray}
\mu^+\mu^-\rightarrow &\tilde{l}^+\tilde{l}^-\  {\rm or}\  
\tilde{\nu}\tilde{\nu}^c&\rightarrow \tau^+\nu+\tilde{\chi}_2^0\tilde{\chi}_1^+
\cr
&&
\left\{\begin{array}{c}
\tilde{\chi}_2^0\rightarrow\tau^+(\rightarrow\e^+) \tau^-(\rightarrow\mu^-)+\tchi_1^0\\
\tchi_1^+\rightarrow 2jets +\tchi_1^0
\end{array}\right.
\end{eqnarray}
is doubly suppressed by $\tau$ branching ratio into 
lepton, and the muon is very soft. The misidentification of 
primary $\tau^+\tau^-$ as primary $\tau\mu$ 
\begin{eqnarray}
\mu^+\mu^-\rightarrow & \tilde{\nu}\tilde{\nu}^c &
\rightarrow \tau^+\tau^-(\rightarrow\mu^-) +  \tchi_1^+ \tchi_1^- 
\cr 
&&
\left\{\begin{array}{c}
\tchi_1^+\rightarrow \e^+\nu_\e +\tchi_1^0\\
\tchi_1^-\rightarrow 2jets+\tchi_1^0
\end{array}\right.
\end{eqnarray}
would be  the dominant background.

\end{itemize}

In the limit of the small LFV, 
the signal  dominates over the background  for the above two process. 
The statistics of the signal will be estimated by
\begin{eqnarray}
S&=& 
{\cal L}\times2 A(p_{\tau}^2 +p_{\tau})  \sigma\left( \tau^+\mu^- \tilde{\chi}^+_1(\rightarrow l^+)
\tilde{\chi}^0_2(\rightarrow 2jets)\right),
\end{eqnarray}
since the background is negligible.
In Section 4.2 we take $p_{\tau}=0.64$
if we use only the energy cut.
Assuming $t$-channel dominance for the production cross section, 
$3\sigma$ discovery of signal corresponds to  
$\sin 2\theta_{\tilde{\nu}} = 0.15$ for 
${\cal L}= 50$ fb$^{-1}$ and  $A= 0.3$.

It is not clear if 
$\tau^+\mu^-\mu^+$   mode is usable. 
The background from sneutrino flavor conserving decay
\begin{eqnarray}
\mu^+\mu^-\rightarrow &\tnu\tnu^c &\rightarrow\mu^+\mu^-\tchi_1^\pm\tchi_1^\mp
\cr
&&
\left\{\begin{array}{c}
\tchi_1^{\pm}\rightarrow \tau^{\pm} \nu(\bar{\nu})+\tilde{\chi}_1^0\\
\tchi_1^{\mp}\rightarrow 2jets+\tilde{\chi}_1^0
\end{array}
\right.
\ (23 {\rm fb} )
\label{B8}
\end{eqnarray}
is large, and would be an important background for this case. 
Both of the $\mu$'s in the background process  have 
energy between $E^{\rm min}_{\mu}$ and $E^{\rm max}_{\mu}$. Therefore the cut 
$E^{\rm soft}_{\mu}< E^{\rm min}_{\mu}<E^{\rm hard}_{\mu}$, 
will remove the background Eq.~(\ref{B8}) completely. 
The cut will remove some signal events,  and the 
signal acceptance  depends on the decay 
distribution of $\tchi_1^{\pm}$. 
The background from $\tau^+\nu\tilde{\chi}^0_2(\rightarrow\mu^+\mu^-)
\tilde{\chi}_1^+(\rightarrow2jets)$
may also  be significant because  
muons come directly from 
the $\tchi_2^0$ decay, unlike the two modes discussed previously, 
while the cross section 
is suppressed only by a  factor of $\sin^2 2 \theta_{\tilde{\nu}}$. 
To remove the background, we 
may have to increase the energy cut for the harder muon. 
The MC simulation is needed to obtain $S/N$ ratio 
with the optimized cuts for this mode. 

\section{Conclusion and Discussion }
In this paper we discuss the detection of the LFV
expected in the MSSMR at future lepton colliders.  We assume 
$\nu_{\tau}$-$\nu_{\mu}$ mixing inspired by the Super-Kamiokande 
atmospheric results. 
In the model, the LFV is induced 
in the left-handed slepton mass matrix radiatively. 
As a result, the charged sleptons and the sneutrinos in the second and 
third generations  mix, and the LFV
may be observed by measuring the slepton production 
and their decay; 
$\mu^+\mu^-(\e^+\e^-) $ $\rightarrow\tnu\tnu^c \rightarrow $ 
$\tau\mu \tchi^+_1\tchi^-_1$ 
and $\mu^+\mu^-(\e^+\e^-)$ $\rightarrow \tl^+\tl^-\rightarrow$ 
$\tau\mu\tchi^0_2\tchi^0_2$
where $ \tchi^{\pm}_1$ and $\tchi^0_2$ decay into leptons or jets with 
the LSP further.

For $\e^+\e^-$ colliders, the signal cross section 
is small because only $s$-channel exchange 
of gauge boson is involved. 
The $\tnu_{\tau}\tnu_{\tau}^c(\tilde{\tau}^+\tilde{\tau}^-)$
production and the decay into $\tau\tau+\tchi_1^+\tchi_1^-
(\tchi^0_2\tchi_2^0)$ would 
be significant background if we do not veto $\mu$ from 
the $\tau$ decay. In this paper, we point out  that 
the background events would  be reduced drastically by 
requiring $E_{\mu}^{\rm cut}\geq E^{\min}_{\mu}$, where 
$E^{\min}_{\mu}$ is the minimum muon energy  for the signal events. 
Furthermore $\sigma_{IP}$, the distance of the muon track 
from the interaction point, can be measured precisely 
for future LC detectors.  Requiring $\sigma_{IP}\ll
c\tau_{\tau}$ to $\tau\mu X$ events we can improve the $S/N$ ratio 
further. For one year of luminosity ${\cal L}=50$fb$^{-1}$, 
$\sin2 \theta_{\tilde{\nu}}\gsim 0.5$ and $\Delta m_{\tilde \nu} \gsim
0.4$GeV  may be explored in the mode $\tau\mu+4jets$. The  experimental 
reach is limited by statistics only. 

At $\mu^+\mu^-$ colliders, 
the signal cross sections are enhanced by the $t$-channel exchange 
of charginos or neutralinos. In the limit of the small LFV, 
the production cross sections for the third generation sfermions 
involve only $s$-channel diagrams, therefore the background 
is relatively suppressed. 
 $\sin2 \theta_{\tilde{\nu}}\gsim 0.1$ and 
$\Delta m_{\tilde {\nu}} \gsim 0.1$GeV may be explored
in the mode $\tau\mu+4jets$ without the $\sigma_{IP}$ cut. 

The $\tau\mu + 4 jets$ mode receives contribution both from 
the $\tnu$ and $\tilde{l}$ productions. The $\tilde{l}$ 
mass matrix is rather complicated,  depending on the 
$\tilde{\tau}_L$-$\tilde{\tau}_R$ mixing. On the other hand,
the $\tau\mu l + 2jets$ 
modes receive the contribution  only from the sneutrino production, 
therefore theoretically clean. We discussed the 
study of the modes at $\mu^+\mu^-$ colliders. Unfortunately, 
the half of the decay modes are masked by SUSY backgrounds.

Because the decay 
of the third generation sleptons
involves $\tau$ leptons and jets, special attention 
must be paid to the $\tau$ isolation cuts. 
No serious MC simulation involving $\tilde{\tau}_L$ and 
$\tilde{\nu}_{\tau}$ has been done so far. In this paper, 
we simply assumed overall acceptance of 30\%. The isolation 
cuts may alter the acceptance by a factor, but we 
expect the problem is less severe for $\tau\mu l + 2 jets$ modes. 

We did not discuss all discovery modes in this paper. For example 
the processes $\tilde{l}\tilde{l}
\rightarrow\tau\mu{\tilde \chi}^0_2(\rightarrow ll) 
{\tilde \chi}^0_2(\rightarrow 2jets)$ 
must be a clean mode sensitive to the LFV in the $\tilde{l}$ mass matrix. 
The branching ratio of the modes  depends on $\tan\beta$ 
sensitively  through $Br(\tilde{\chi}^0_2\rightarrow 2l)$. We 
also did not discuss the case where $\tilde{l}$ decays dominantly 
into $l{\tilde \chi}^0_1$ since  the region 
where $m_{\tilde{l}_L}$, $m_{\tilde{\nu}} < m_{\tilde{\chi}_1^-}$, 
$m_{\tilde{\chi}_2^0}$ is expected to be very narrow 
in the minimal supergravity scenario.
In the case, the signal 
is $\tau\mu+\E$ . The mode suffers background from 
$W$ pair production unless high beam polarization 
is available in LC's.  In $\mu^+\mu^-$ colliders the signal may be 
accessible due to the high statistics. We did not discuss $\tau$-$\e$ mixing,
which is not favored by current experimental data for neutrinos. 

Atmospheric neutrino study implies that the Yukawa sector of 
lepton could be completely different from 
that of quark. Namely large  mixing 
is expected between the second and  third generations
for lepton, while it is very small for quark 
mass matrix.  
If this 
implies existence of the large LFV Yukawa interaction between the 
third and second generations, it leads to the LFV signal 
in the slepton production in the seesaw mechanism. 
We emphasize that a $\mu^+\mu^-$ collider stands 
as a powerful tool to explore it in future. 

\underline{Acknowledgment}

The authors would like to thank Dr. Y.~Okada, Dr. K.~Fujii, 
Dr. A.~Miyamoto, and Dr. M.~Drees.
This work is supported in part by  Grants in aid for Science and 
Culture of Japan (No.10740133, No.10140211, and No.10140216).

\newpage
\appendix

\section{Lagrangian of the MSSM with Flavor Violations}
 
In this Appendix, we present the Lagrangian of the MSSM.
The superpotential in the MSSM is given by
\begin{eqnarray}
\label{superpotentialMSSM}
W&=&f_l^{ij} H_1 E_i^c  L_j
+ f_{d}^{ij} H_1 D_i^c  Q_j
+ f_{u}^{ij} H_2 U_i^c  Q_j
+ \mu H_1 H_2
\end{eqnarray}
where $L_i$ represents the chiral multiplet of an $SU(2)_L$ doublet
lepton, $E_i^c$ an $SU(2)_L$ singlet charged lepton, $H_1$ and $H_2$ two 
Higgs doublets with opposite hypercharge.
Similarly $Q$, $U^c$ and $D^c$ represent chiral multiplets of quarks of an
$SU(2)_L$ doublet and two singlets with different $U(1)_Y$ charges.
The subscripts $i$ and $j$ are for generation. In this article we ignore
the CKM mixing for quark sector, which is irrelevant for our discussion.

The soft SUSY breaking terms are given as
\begin{eqnarray}
\label{softbreaking}
-{\cal{L}}_{soft}&=&(m_{\tilde Q}^2)_i^j {\tilde q}_{L}^{\dagger i}
{\tilde q}_{Lj}
+(m_{\tilde u}^2)^i_j {\tilde u}_{Ri}^* {\tilde u}_{R}^j
+(m_{\tilde d}^2)^i_j {\tilde d}_{Ri}^* {\tilde d}_{R}^j
\nonumber \\
& &+(m_{\tilde L}^2)_i^j {\tilde l}_{L}^{\dagger i}{\tilde l}_{Lj}
+(m_{\tilde e}^2)^i_j {\tilde e}_{Ri}^* {\tilde e}_{R}^j
+{\tilde m}^2_{h1}h_1^{\dagger} h_1
+{\tilde m}^2_{h2}h_2^{\dagger} h_2
\nonumber \\
& &+(B \mu h_1 h_2 +h.c.)
\nonumber \\
& &+ ( A_d^{ij}h_1 {\tilde d}_{Ri}^*{\tilde q}_{Lj}
+A_u^{ij}h_2 {\tilde u}_{Ri}^*{\tilde q}_{Lj} 
+A_l^{ij}h_1 {\tilde e}_{Ri}^*{\tilde l}_{Lj}
+h.c.)
\nonumber \\
& & 
+\frac{1}{2}M_1 {\tilde B} {\tilde B}
+\frac{1}{2}M_2 {\tilde W} {\tilde W}
+\frac{1}{2}M_3 {\tilde G} {\tilde G} + h.c.).
\end{eqnarray}
Here, the first two lines are  soft terms for  sleptons, squarks, 
and the Higgs bosons, and the third and fourth lines are  those
for a supersymmetric mass and Yukawa interactions,
while the last line gives gaugino mass terms.

\section{Interaction of SUSY particles}

In this Appendix, we give our notations and conventions for masses and 
vertices relevant for our calculation. First, we discuss quarks and leptons.  
We denote by $l_i$, $u_i$, and $d_i$
the fermion mass eigenstates with mass $m_{f_i}$ ($f=l,u,d$).  
As for the neutrinos, their masses are small and negligible.  In our
convention, $\nu_i$ is 
the $SU(2)_L$ isodoublet partner to $e_{Li}$.

Next, we consider sfermions. Let $\tilde f_{Li}$ and $\tilde f_{Ri}$ be the
superpartners of $f_{Li}$ and $f_{Ri}$, respectively.  Here, $f$ stands 
for $l$, $u$, or $d$.  The sfermion mass matrix 
can be written in the following form,
\begin{equation}
     \left ( \tilde f^{\dagger}_{L}, \tilde f^{\dagger}_R \right )
     \left ( \begin{array}{cc}
               m_L^2    & m_{LR}^{2 {\sf T}} \\
               m_{LR}^{2} & m_R^2   
             \end{array}                  \right )
     \left (  \begin{array}{c}
              \tilde f_{L} \\ \tilde f_R 
              \end{array}                 \right ),
\end{equation}
where $m_L^2$ and $m_R^2$ are $(3 \times 3)$ hermitian matrices and 
$m_{LR}^2$ is a $3\times 3$  matrix. These elements are given from 
Eqs.~(\ref{superpotentialMSSM},\ref{softbreaking}) as following,
\begin{eqnarray}
m_L^2 &=& 
m_{\tilde f_L}^2 + m_f^2 
+ m_Z^2 c_{2 \beta} (T_{3L}^f -Q_{em}^f s_W^2),
\\
m_R^2 &=& 
m_{\tilde f_R}^2 
+ m_f^2 + m_Z^2 c_{2 \beta} Q_{em}^f s_W^2,
\\
m_{LR}^2 &=& 
\left\{
\begin{array}{c}
-A_f v s_\beta /\sqrt{2}  - m_f \mu/t_\beta ~~~~ (f=u),
\\
A_f v c_\beta /\sqrt{2} - m_f \mu t_\beta ~~~~ (f=d,l),
\end{array}
\right.
\end{eqnarray}
where $T_{3L}^f$ and $Q^f_{em}$ are the weak isospin and the electric charge, 
respectively. Here, $m_{\tilde f_L}^2 = m_{\tilde Q}^2$ for squarks,
$m_{\tilde f_L}^2 = m_{\tilde L}^2$ for sleptons, and 
$m_{\tilde f_R}^2$ is each right-handed sfermion soft-breaking 
mass. The vacuum expectation values for the Higgs doublets are given 
as $\langle h_1 \rangle = (v \cos\beta,0)^T$ and 
$\langle h_2 \rangle = (0,v \sin\beta)^T$.\footnote{
For simplicity, we denote  $\sin\beta$ and $\cos\beta$ as 
$s_\beta$ and $c_\beta$ in equations. Similarly, 
$s_W$ and $c_W$ are for the Weinberg angle, $\sin\theta_W$ and $\cos\theta_W$.
}
We assume the above mass matrix to be real.
This is, in general, not diagonal, and includes mixing between
different generations.  For charged sleptons and squarks, 
we diagonalize the mass matrix  
$M_{\tilde{f}}^2$ by  a $(6\times 6)$ real orthogonal matrix $U^f$ as
\begin{equation}
    U^f M_{\tilde f}^2 U^{f {\sf T}} =({\rm diagonal}),
\end{equation}
and we denote its eigenvalues by $m^2_{\tilde f_X}$ ($X= 1, \cdots,
6$).  The mass eigenstates are then written as
\begin{equation}
   \tilde f_X = U^f_{X,i} \tilde f_{Li} + U^f_{X,i+3} \tilde f_{Ri},
   \hspace{1.5cm} (X=1, \cdots, 6).
\end{equation}
Conversely, we have
\begin{eqnarray}
   \tilde f_{Li} = &U^{f{\sf T}}_{iX} \tilde f_X & =U^{f}_{Xi} \tilde f_X,
\\
   \tilde f_{Ri} =& U^{f{\sf T}}_{i+3,X} \tilde f_X & = U^f_{X,i+3} \tilde
   f_X.
\end{eqnarray}
An attention should be paid to the  neutrinos since the MSSM has no
right-handed sneutrino.  Let $\tilde \nu_{Li}$ be the
superpartner of the neutrino $\nu_{i}$.  The mass eigenstate
$\tilde \nu_X$ ($X=1,2,3$) is related to  $\tilde \nu_{Li}$ as
\begin{equation}
   \tilde \nu_{Li}=U^{\nu}_{Xi} \tilde \nu_{X}.
\end{equation}

We now turn to charginos.  The chargino mass matrix is given by
\begin{equation}
  -{\cal L}_m =
     \left ( \overline{\tilde W^-_R}~ \overline{\tilde H^-_{2R}} \right )
     \left (  \begin{array}{cc}
                M_2            & \sqrt{2} m_W c_\beta \\
                \sqrt{2}m_W s_\beta &  \mu              
              \end{array}                                 \right)  
     \left (  \begin{array}{c}
              \tilde W^-_L    \\ \tilde H_{1L}^-
              \end{array}                                 \right) +h.c..
\end{equation}
This matrix $M_C$ is diagonalized by $(2 \times 2)$ real orthogonal matrices
$O_L$ and $O_R$ as
\begin{equation}
     O_R M_C O_L^{\sf T} =({\rm diagonal}).
\end{equation}
Define
\begin{equation}
   \left( \begin{array}{c} 
           \tilde \chi^-_{1L} \\  
           \tilde \chi^-_{2L}
           \end{array}                 \right)
  =O_L   \left( \begin{array}{c}
                \tilde W^-_L  \\
                \tilde H^-_{1L}
                \end{array}            \right),
\hspace{1.5cm}
    \left( \begin{array}{c} 
           \tilde \chi^-_{1R} \\  
           \tilde \chi^-_{2R}
           \end{array}                 \right)
  =O_R   \left( \begin{array}{c}
                \tilde W^-_R  \\
                \tilde H^-_{2R}
                \end{array}            \right).
\end{equation}
Then 
\begin{equation}
    \tilde \chi^-_A  =\tilde \chi^-_{AL} + \tilde \chi^-_{AR},
\hspace{1.5cm}
    (A=1,2),
\end{equation}
forms a Dirac fermion with the mass $M_{\tilde \chi^-_A}$.

Finally we consider neutralinos.  The neutralino mass matrix 
is given by
\begin{equation}
 -{\cal L}_m =
  \frac{1}{2}  
  \left( \tilde B_L \tilde W^0_L \tilde H^0_{1L} \tilde H^0_{2L}
  \right) 
   M_N 
   \left(   \begin{array}{c}
            \tilde B_L  \\
            \tilde W^0_L \\
            \tilde H^0_{1L} \\
            \tilde H^0_{2L}
            \end{array}                  \right)  +h.c.,
\end{equation}
where 
\begin{equation}
   M_N=
  \left(
   \begin{array}{cccc}
     M_1    & 0 & -m_Z s_W c_\beta & m_Z s_W s_\beta \\
     0 & M_2 & m_Z c_W c_\beta & -m_Zc_W s_\beta \\
     -m_Z s_W c_\beta & m_Z c_W c_\beta & 0 & -\mu
     \\ 
     m_Z s_W s_\beta & -m_Z c_W s_\beta & -\mu & 0 
   \end{array}            \right).
\end{equation}
The diagonalization is done by a real orthogonal matrix $O_N$, 
\begin{equation}
    O_N M_N O_N^{\sf T} = ({\rm diagonal}).
\end{equation}
The mass eigenstates are given by
\begin{equation}
         \tilde \chi^0_{AL} =(O_N)_{AB} \tilde X^0_{BL},
\hspace{1.5cm}
           (A,B=1, \cdots ,4),
\end{equation}
where
\begin{equation}
   \tilde X^0_{AL} = ( \tilde B_L, \tilde W^0_L, \tilde H^0_{1L},
   \tilde H^0_{2L}).
\end{equation}
We have thus Majorana spinors
\begin{equation}
   \tilde \chi^0_A = \tilde \chi^0_{AL} + \tilde \chi^0_{AR},
\end{equation}
 with the mass $M_{\tilde \chi^0_A}$.

We now give the interaction Lagrangian of the SUSY particles. First, 
the fermion-sfermion-chargino interaction is given as 
\begin{eqnarray}
   {\cal L}_{\rm int}& = & 
     -g_2 \overline{\tilde \chi^-_A}  (C^{R (l)}_{iAX} P_R+ C^{L(l)}_{iAX} P_L ) 
     l_i \tilde \nu_X^\dagger 
\nonumber \\
  & &- g_2 \overline{\tilde \chi^+_A}  (C^{R (\nu)}_{iAX} P_R+ C^{L(\nu)}_{iAX} P_L )
     \nu_i \tilde l_X^\dagger
\nonumber \\
  & &- g_2 \overline{\tilde \chi^-_A} (C^{R (d)}_{iAX} P_R+ C^{L(d)}_{iAX} P_L ) 
     d_i \tilde u_X^\dagger
\nonumber \\
  & &- g_2 \overline{\tilde \chi^+_A} (C^{R (u)}_{iAX} P_R+ C^{L(u)}_{iAX} P_L ) 
     u_i \tilde d_X^\dagger
   +h.c.,
\end{eqnarray}
where the coefficients are
\begin{eqnarray}
 C^{L(l)}_{iAX}& =& (O_R)_{A1} U^{\nu}_{X,i},
\nonumber \\
 C^{R(l)}_{iAX}& = & -\frac{m_{l_i}}{\sqrt{2}m_W c_\beta}(O_L)_{A2}
                    U^{\nu}_{X,i},
\nonumber \\
 C^{L(\nu)}_{iAX} &= & \{(O_L)_{A1} U^l_{X,i} 
                       - \frac{m_{l_i}}{\sqrt{2}m_W c_\beta}(O_L)_{A2}
                     U^l_{X,i+3} \},
\nonumber \\
  C^{R(\nu)}_{iAX} &=&0,
\nonumber \\
 C^{L(d)}_{iAX} & = & \{ (O_R)_{A1} U^u_{X,i}
-\frac{m_{u_i}}{\sqrt{2}m_W s_\beta} (O_R)_{A2} U^u_{X,i+3} \},
\nonumber \\
 C^{R(d)}_{iAX}& = & -\frac{m_{d_i}}{\sqrt{2}m_W c_\beta}(O_L)_{A2}
                    U^u_{X,i},
\nonumber \\
 C^{L(u)}_{iAX} & = & \{ (O_L)_{A1}U^d_{X,i}
-\frac{m_{d_i}}{\sqrt{2}m_W c_\beta}(O_L)_{A2} U^d_{X,i+3} \},
\nonumber \\
 C^{R(u)}_{iAX} & = & -\frac{m_{u_i}}{\sqrt{2}m_W s_\beta}
                    (O_R)_{A2} U^d_{X,i}.
\end{eqnarray}
The interaction Lagrangian of fermion-sfermion-neutralino is similarly 
written as
\begin{equation}
  {\cal L}_{\rm int}
=  -g_2 \overline{\tilde \chi^0_A} (N^{R(f)}_{iAX} P_R +N^{L(f)}_{iAX} P_L) f_i 
    \tilde f_X^\dagger +h.c.,
\end{equation}
where $f$ stands for $l,\nu,d$, and $u$.  The coefficients
 are
\begin{eqnarray}
  N^{L(l)}_{iAX}&=& \frac{1}{\sqrt{2}} \{
       [-(O_N)_{A2} -(O_N)_{A1} t_W] U^l_{X,i}
        + \frac{m_{l_i}}{m_Wc_\beta} (O_N)_{A3} U^l_{X,i+3} \},
\nonumber \\
  N^{R(l)}_{iAX} &=& \frac{1}{\sqrt{2}} \{ 
           \frac{m_{l_i}}{m_W c_\beta} (O_N)_{A3} U^{l}_{X,i} 
           +2 (O_N)_{A1} t_W U^{l} _{X,i+3} \},
\nonumber \\
  N^{L(\nu)}_{iAX} &=& \frac{1}{\sqrt{2}}
             [(O_N)_{A2}-(O_N)_{A1} t_W] U^{\nu}_{X,i},
\nonumber \\
  N^{R(\nu)}_{iAX}&=&0,
\nonumber \\
  N^{L(d)}_{iAX}&=& \frac{1}{\sqrt{2}} \{
      [-(O_N)_{A2}+\frac{1}{3}(O_N)_{A1} t_W] U^d_{X,i}
     +\frac{m_{d_{i}}}{m_W c_\beta}(O_N)_{A3} U^d_{X,i+3} \},
\nonumber \\
  N^{R(d)}_{iAX}&=& \frac{1}{\sqrt{2}}
     \{ \frac{m_{d_{i}}}{m_W c_\beta}(O_N)_{A3} U^d_{X,i}
       +\frac{2}{3} t_W (O_N)_{A1} U^d_{X,i+3} \},
\nonumber \\
  N^{L(u)}_{iAX}&=& \frac{1}{\sqrt{2}} 
         \{ [(O_N)_{A2} +\frac{1}{3} (O_N)_{A1} t_W]
         U^u_{X,i}
        +\frac{m_{u_i}}{m_W s_\beta} (O_N)_{A4} U^u_{X,i+3} \},
\nonumber \\
  N^{R(u)}_{iAX}&=& \frac{1}{\sqrt{2}}
        \{ \frac{m_{u_i}}{m_Ws_\beta} (O_N)_{A4} U^u_{X,i}
         -\frac{4}{3} t_W (O_N)_{A1} U^u_{X,i+3}  \}.
\end{eqnarray}

Next, the interaction of sfermion-$Z$ boson is presented as 
\begin{eqnarray}
{\cal L}&=& -i g_Z z^{(\tilde{f})}_{XY} \tilde{f}_X^\dagger {\partial}^\mu \tilde{f}_Y Z_\mu +h.c.,
\nonumber
\end{eqnarray}
where
\begin{eqnarray}
z^{(\tilde{f})}_{XY}&=&
T_{3L}^f \sum^3_{i=1} U^f_{X,i} U^f_{Y,i} - Q s_W^2 \delta_{XY},
\end{eqnarray}
and $g_Z^2=g_2^2+g_Y^2$.
The interaction of chargino or neutralino to $Z$ boson 
is given as
\begin{eqnarray}
{\cal L} &=& - \frac12 g_Z z^{(\tilde \chi^0)}_{AB} 
\overline{\tilde \chi^0_A} \gamma^\mu P_L {\tilde \chi^0_B} Z_\mu
\nonumber\\
&&-g_Z z^{(\tilde \chi^-)}_{AB}
\overline{\tilde \chi^-_A} \gamma^\mu 
\left(z^{(\tilde \chi^-)}_{LAB} P_L+z^{(\tilde \chi^-)}_{RAB} P_R\right) 
{\tilde \chi^-_B} Z_\mu,
\end{eqnarray}
where
\begin{eqnarray}
z^{(\tilde \chi^0 )}_{AB} &=&
\left((O_N)_{A3}(O_N)_{B3}-(O_N)_{A4}(O_N)_{B4}\right),
\nonumber\\
z^{(\tilde \chi^- )}_{LAB} &=&
\left(\frac12 (O_L)_{A2}(O_L)_{B2} - c_W^2 \delta_{AB}\right),
\nonumber\\
z^{(\tilde \chi^- )}_{RAB} &=&
\left(\frac12 (O_R)_{A2}(O_R)_{B2} - c_W^2 \delta_{AB}\right).
\end{eqnarray}
The interaction of chargino and neutralino to $W$ boson 
is also 
\begin{eqnarray}
{\cal L} &=& 
-g_2 \overline{\tilde \chi^0_A} \gamma_\mu 
\left(w^{(\tilde{\chi})}_{LAB} P_L +w^{(\tilde{\chi})}_{RAB} P_R\right)
{\tilde \chi^-_B} W_\mu^+ + h.c.,
\end{eqnarray}
where
\begin{eqnarray}
w^{(\tilde{\chi})}_{LAB} &=&
\left((O_N)_{A2}(O_L)_{B1}+\frac1{\sqrt{2}}(O_N)_{A3}(O_L)_{B2}\right),
\nonumber\\
w^{(\tilde{\chi})}_{RAB} &=&
\left((O_N)_{A2}(O_R)_{B1}-\frac1{\sqrt{2}}(O_N)_{A4}(O_R)_{B2}\right).
\end{eqnarray}

\section{Matrix elements of the off-shell slepton production and 
decay processes}

We present the matrix elements of the off-shell slepton 
production and decay processes at $l_i^+ l_i^-$ collisions.  First, we discuss 
the charged slepton pair production process 
with the center of mass energy $\sqrt{s}$. The kinematics for the process  
($\l_i^-(p) ~\l_i^+(\bar p) \rightarrow \tilde{l}_X^- (k)~\tilde{l}_Y^+
(\bar k)$) is given as 
\begin{eqnarray}
     p &=&\frac{\sqrt{s}}{2} (1,0,0,1), \nonumber\\
\bar p &=&\frac{\sqrt{s}}{2} (1,0,0,-1), \nonumber\\
     k &=&(\frac{s+\Delta k^2}{2\sqrt{s}}, 
           \frac{\sqrt{s}}{2} \beta_p s_\theta, 0,
           \frac{\sqrt{s}}{2} \beta_p c_\theta),
        \nonumber\\
\bar k &=&(\frac{s-\Delta k^2}{2\sqrt{s}}, 
           -\frac{\sqrt{s}}{2} \beta_p s_\theta, 0,
           -\frac{\sqrt{s}}{2} \beta_p c_\theta),
\end{eqnarray}
where
\begin{eqnarray}
&&\Delta k^2 = k^2-\bar{k}^2,
\nonumber\\
\beta_p^2 &=&\frac{1}{s^2}\left({s^2-2 s(k^2+\bar{k}^2)
+(\Delta k^2)^2}\right).
\nonumber
\end{eqnarray}
The amplitudes ${\cal M}_{XY}^{(h\bar{h})}$ with the helicity of
$l_i^-(l_i^+)$ being $h(\bar{h})$ are given as 
\begin{eqnarray}
{\cal M}^{(+-)}_{XY}(k^2,\bar{k}^2) &=& - s \beta_p s_\theta
\left\{
  \frac{e^2}{s} \delta_{XY} 
+ \frac{g_Z^2}{s-m_Z^2} z^{(\tilde{l})}_{XY} z^{(l)}_R
+ \sum_A \frac12 \frac{g_2^2} {t-M_{\tilde{\chi}^0_A}^2}N^{R(l)}_{iAX} N^{R(l)}_{iAY}
\right\},
\nonumber\\
{\cal M}^{(-+)}_{XY}(k^2,\bar{k}^2) &=& s \beta_p s_\theta
\left\{
  \frac{e^2}{s} \delta_{XY} 
+ \frac{g_Z^2}{s-m_Z^2} z^{(\tilde{l})}_{XY} z^{(l)}_L
+ \sum_A \frac12 \frac{g_2^2} {t-M_{\tilde{\chi}^0_A}^2} N^{L(l)}_{iAX} N^{L(l)}_{iAY}
\right\},
\nonumber\\
{\cal M}^{(++)}_{XY}(k^2,\bar{k}^2) &=& - \sum_A \sqrt{s} 
\frac{g_2^2} {t-M_{\tilde{\chi}^0_A}^2} N^{R(l)}_{iAX} N^{L(l)}_{iAY}M_{\tilde{\chi}^0_A},
\nonumber\\
{\cal M}^{(--)}_{XY}(k^2,\bar{k}^2) &=& - \sum_A \sqrt{s } 
\frac{g_2^2} {t-M_{\tilde{\chi}^0_A}^2} N^{L(l)}_{iAX} N^{R(l)}_{iAY} M_{\tilde{\chi}^0_A},
\nonumber
\end{eqnarray}
where $z^{(l)}_L (z^{(l)}_R)=-1/2+s_W^2 (s_W^2)$ and 
$t = (\Delta k^2)^2/4s - s(1+\beta_p^2 -2 \beta_p c_\theta)/4$. Form these
amplitudes the cross section of the on-shell charged slepton pair production 
($\l_i^- ~\l_i^+ \rightarrow \tilde{l}_X^-~\tilde{l}_Y^+$)
is given as
\begin{equation}
\frac{d\sigma}{d c_\theta} = 
\frac1{2s}\frac{\beta_p}{16\pi} \overline{\sum_{h\bar h}}\left|{\cal M}^{(h\bar h)}_{XY}
(m^2_{\tilde{l}_X},m^2_{\tilde{l}_Y}) \right|^2,
\end{equation}
where $\overline{\Sigma}_{h\bar h}$ means the average over the helicities 
of the initial beams.
Similarly, the helicity amplitudes of off-shell sneutrino pair production
($\l_i^-(p) ~\l_i^+(\bar p) \rightarrow \tilde{\nu}_X (k)~\tilde{\nu}_Y
(\bar k)$) are given as 
\begin{eqnarray}
{\cal M}^{(+-)}_{XY}(k^2,\bar{k}^2) &=& - s \beta_p s_\theta
\left\{
\frac{g_Z^2}{s-m_Z^2} z^{(\tilde{\nu})}_{XY} z^{(l)}_R
+ \sum_A \frac12 \frac{g_2^2} {t-M_{\tilde{\chi}^-_A}^2}C^{R(l)}_{iAX} C^{R(l)}_{iAY}
\right\},
\nonumber\\
{\cal M}^{(-+)}_{XY}(k^2,\bar{k}^2) &=& s \beta_p s_\theta
\left\{
\frac{g_Z^2}{s-m_Z^2} z^{(\tilde{\nu})}_{XY} z^{(l)}_L
+ \sum_A \frac12 \frac{g_2^2} {t-M_{\tilde{\chi}^-_A}^2} C^{L(l)}_{iAX} C^{L(l)}_{iAY}
\right\},
\nonumber\\
{\cal M}^{(++)}_{XY}(k^2,\bar{k}^2) &=& - \sum_A \sqrt{s} 
\frac{g_2^2} {t-M_{\tilde{\chi}^-_A}^2} C^{R(l)}_{iAX} C^{L(l)}_{iAY}M_{\tilde{\chi}^-_A},
\nonumber\\
{\cal M}^{(--)}_{XY}(k^2,\bar{k}^2) &=& - \sum_A \sqrt{s} 
\frac{g_2^2} {t-M_{\tilde{\chi}^-_A}^2} C^{L(l)}_{iAX} C^{R(l)}_{iAY} M_{\tilde{\chi}^-_A}.
\nonumber
\end{eqnarray}

Next, we show the matrix elements for the decay processes of the off-shell 
slepton.  Now we assume that sleptons decay to two-body states. 
Those for $\tilde{l}_X^-\rightarrow \tilde{\chi}_A^-~\nu_i$  
(${\cal M}_{XAi}$) are
\begin{eqnarray}
{\cal M}_{XAi}(k^2) &=& g_2 C^{L (\nu)}_{iAX} \sqrt{k^2 \beta_d},
\end{eqnarray}
where $k^2$ is square of the momentum of  the slepton and 
$\beta_d = 1-M_{\tilde{\chi}_A^-}^2/k^2$. We ignore the lepton mass.
The partial decay width of the on-shell slepton is 
\begin{eqnarray}
\Gamma
&=&
\frac{1}{16 \pi m_{\tilde{l}_X}} \beta_d 
\left|{\cal M}_{XAi}(m_{\tilde{l}_X})\right|^2.
\end{eqnarray}
The matrix elements for $\tilde{l}_X\rightarrow \tilde{\chi}_A^0~l_i$ 
with the helicity of $l_i$ being $h$ (${\cal M}_{XAi}^{(h)}$) are 
\begin{eqnarray}
{\cal M}_{XAi}^{(+)}(k^2) &=& g_2 N^{R (l)}_{iAX} \sqrt{k^2\beta_d}, 
\nonumber\\
{\cal M}_{XAi}^{(-)}(k^2) &=& g_2 N^{L (l)}_{iAX} \sqrt{k^2\beta_d}. 
\end{eqnarray}
Similarly, the matrix elements for $\tilde{\nu}_X\rightarrow \tilde{\chi}_A^+~ 
l_i$ with the helicity of $l_i$ being $h$ (${\cal M}_{XAi}^{(h)}$) are
\begin{eqnarray}
{\cal M}_{XAi}^{(+)}(k^2) &=& g_2 C^{R (l)}_{iAX} \sqrt{k^2\beta_d}, 
\nonumber\\
{\cal M}_{XAi}^{(-)}(k^2) &=& g_2 C^{L (l)}_{iAX} \sqrt{k^2\beta_d}, 
\end{eqnarray}
and those for $\tilde{\nu}_X\rightarrow \tilde{\chi}_A^0~\nu_i$ are
\begin{eqnarray}
{\cal M}_{XAi}(k^2) &=& g_2 N^{L (\nu)}_{iAX} \sqrt{k^2 \beta_d}.
\end{eqnarray}

\section{Partial decay widths for charginos and neutralinos}

In this Appendix, we present the partial decay widths of charginos and
neutralinos to three-body states. The partial decay width of
chargino into neutralino is
\begin{eqnarray}
\Gamma(\tilde{\chi}^-_A\rightarrow \tilde{\chi}^0_B 
\bar{f}_{\uparrow i} f_{\downarrow j}) 
&=&
N_C \frac{1}{256 \pi^3}
\int |{\cal M}|^2 M_{\tilde{\chi}^-_A}d x d y,
\end{eqnarray}
where $N_C$ is a color factor 
($N_C=3$ for $(\bar{f}_{\uparrow i}, f_{\downarrow j})=(\bar u, d)$ and
$N_C=1$ for $(\bar{f}_{\uparrow i}, f_{\downarrow j})=(\bar\nu, l)$).
The squared amplitude is 
\begin{eqnarray}
|{\cal M}|^2 &=& 
2 A_L^2 (1-y) (y-r_{\tilde{\chi}_B}^2) 
+2 A_R^2 (1-x) (x-r_{\tilde{\chi}_B}^2) 
\nonumber\\
&&-4 A_L A_R r_{\tilde{\chi}_B} z,
\end{eqnarray}
and the coefficients are given as
\begin{eqnarray}
A_L&=&\frac{g_2^2}{\sqrt2} \left\{ 
\frac{w^{(\tilde{\chi})}_{LBA}} {z-r_W^2}
-\sum_X \frac1{\sqrt2}\frac{C^{L(f_\uparrow)}_{iAX} N^{L(f_\downarrow)}_{jBX}}{y-r_{\tilde{f}_{\downarrow X}}^2}
\right\},
\nonumber\\
A_R&=&\frac{g_2^2}{\sqrt2} \left\{ 
\frac{w^{(\tilde{\chi})}_{RBA}} {z-r_W^2}
+\sum_X \frac1{\sqrt2}\frac{C^{L(f_\downarrow)}_{jAX} N^{L(f_\uparrow)}_{iBX}}{x-r_{\tilde{f}_{\uparrow X}}^2}
\right\}.
\nonumber
\end{eqnarray}
Here, we ignore the Yukawa interaction of Higgsinos. 
The mass ratios $r_{\tilde{\chi}_B}$, $r_{W}$, $r_{\tilde{f}_{\uparrow X}}$, 
and $r_{\tilde{f}_{\downarrow X}}$ are defined as 
\begin{eqnarray}
r_{\tilde{\chi}_B} = \frac{M_{\tilde{\chi}^0_B}}{M_{\tilde{\chi}^-_A}},
&&
r_{W} = \frac{m_W}{M_{\tilde{\chi}^-_A}},
\nonumber\\
r_{\tilde{f}_{\uparrow X}} = \frac{m_{\tilde{f}_{\uparrow X}}}{M_{\tilde{\chi}^-_A}},
&&
r_{\tilde{f}_{\downarrow X}} = \frac{m_{\tilde{f}_{\downarrow X}}}{M_{\tilde{\chi}^-_A}}.
\nonumber
\end{eqnarray}
The boundary condition of the phase space is given as 
\begin{eqnarray}
&z (xy-r_{\tilde{\chi}_B}^2) \ge 0,&
\nonumber\\
&r_{\tilde{\chi}_B}^2 \le x,y \le 1,&
\label{psd}
\end{eqnarray}
and  $x+y+z =1+r_{\tilde{\chi}_B}^2$. 

The partial decay width of neutralino into the lighter neutralino is given as 
\begin{eqnarray}
\Gamma(\tilde{\chi}^0_A\rightarrow \tilde{\chi}^0_B 
\bar{f}_i f_j) 
&=&
N_C \frac{1}{256 \pi^3}
\int |{\cal M}|^2 M_{\tilde{\chi}^0_A}d x d y,
\end{eqnarray}
The squared amplitude is 
\begin{eqnarray}
|{\cal M}|^2 &=& 
2 (A_{LL}^2+A_{RR}^2) (1-y) (y-r_{\tilde{\chi}_B}^2) 
+2 (A_{LR}^2+A_{RL}^2) (1-x) (x-r_{\tilde{\chi}_B}^2) 
\nonumber\\
&&-4 (A_{LL}A_{RL}+A_{RR}A_{LR}) r_{\tilde{\chi}_B} z,
\end{eqnarray}
and the coefficients are given as
\begin{eqnarray}
A_{LL} &=&  \frac12 g_Z^2 \frac{z^{(\tilde \chi^0)}_{BA} z^{(f)}_L}{z-r_Z^2} 
           -\sum_X \frac12 g_2^2 \frac{N^{L(f)}_{jAX} N^{L(f)}_{iBX}}
                              {y-r_{\tilde{f}_X}^2},
\nonumber\\
A_{RL} &=& -\frac12 g_Z^2 \frac{z^{(\tilde \chi^0)}_{BA} z^{(f)}_L}{z-r_Z^2} 
           +\sum_X \frac12 g_2^2 \frac{N^{L(f)}_{iAX} N^{L(f)}_{jBX}}
                              {x-r_{\tilde{f}_X}^2},
\nonumber\\
A_{LR} &=&  \frac12 g_Z^2 \frac{z^{(\tilde \chi^0)}_{BA} z^{(f)}_R}{z-r_Z^2} 
           +\sum_X \frac12 g_2^2 \frac{N^{R(f)}_{iAX} N^{R(f)}_{jBX}}
                              {x-r_{\tilde{f}_X}^2},
\nonumber\\
A_{RR} &=& -\frac12 g_Z^2 \frac{z^{(\tilde \chi^0)}_{BA} z^{(f)}_R}{z-r_Z^2} 
           -\sum_X \frac12 g_2^2 \frac{N^{R(f)}_{jAX} N^{R(f)}_{iBX}}
                              {y-r_{\tilde{f}_X}^2},
\nonumber
\end{eqnarray}
where $z^{(f)}_L (z^{(f)}_R)= T_{3L}  - Q s_W^2 (- Q s_W^2)$.
Here, we also ignore the Yukawa interaction. The mass ratios 
$r_{\tilde{\chi}_B}$, $r_{Z}$, and $r_{\tilde{f}_X}$ 
are defined as 
\begin{eqnarray}
r_{\tilde{\chi}_B} = \frac{M_{\tilde{\chi}^0_B}}{M_{\tilde{\chi}^0_A}},~
r_{Z} = \frac{m_Z}{M_{\tilde{\chi}^-_A}},~
r_{\tilde{f}_X} = \frac{m_{\tilde{f}_X}}{M_{\tilde{\chi}^-_A}}.
\nonumber
\end{eqnarray}
The boundary condition of the phase space is the same as that in 
Eq.~(\ref{psd}).

\section{Energy and IP distribution of muon from tau decay in flight}

In this Appendix we present the energy and the IP distribution of 
muons from  decays of tau leptons in flight with the fixed energy $E_\tau$.
Here, we define $\sigma_{IP}$
as distance between the interaction point where the tau lepton is produced
and the track of muon.

First, we show the angular distribution of muon from tau lepton in flight.
The momentum of tau lepton is given as $E_\tau(1,0,0,1)$, and that of muon is 
$E_\mu(1,s_\theta c_\phi,s_\theta s_\phi,c_\theta)$. In this basis, 
the angular distribution is 
\begin{eqnarray} 
\frac{d\Gamma}{\Gamma}
&=&
\rho(\gamma_{\tau}\theta,z)
dz d (\gamma_{\tau}\theta), 
\end{eqnarray}
where $z=E_\mu/E_\tau$, $\gamma_{\tau}=E_\tau/m_\tau$.
The distribution function $\rho(y,z)$ is 
\begin{equation} 
\rho(y,z)
=
4z^2 y\left(
 3(1+y^2)
-2 z(1+y^2)^2 
+P_\tau \left\{
 (1-y^2)
-2z (1-y^4) 
\right\}
\right)
\end{equation}
where $P_\tau$ is the helicity of tau lepton. Here, we take a limit of 
$\gamma_{\tau}^{-1}\rightarrow 0$. The boundary condition is  
\begin{eqnarray}
&0\le (\gamma_{\tau}\theta)^2 \le 1/z-1,&
\nonumber\\
&0\le z\le1.&
\end{eqnarray}
By integrating $\theta$, the energy distribution of 
muon from decay of tau lepton in flight is given as
\begin{equation} 
\frac{d\Gamma}{\Gamma}
=
\frac13 (1-z) (5+5z-4z^2+P_\tau(1+z-8z^2)) dz.
\end{equation}

Since $\sigma_{IP}$ is defined as distance 
between the interaction point and the  track of muon, 
the IP parameter is 
\begin{equation}
\sigma_{IP}\equiv
\beta_{\tau}\gamma_{\tau}  c t s_\theta
\simeq c(\gamma_{\tau}  \theta) t,
\end{equation}
where $t$ is time between production and decay of $\tau$
and $\beta^2_\tau=1-1/\gamma_\tau^2$.
Then, the energy and the IP distribution is
\begin{eqnarray}
\frac{d\Gamma}{\Gamma}
&=&
\int^\infty_0 \frac{dt}{\tau_\tau} \e^{-\frac{t}{\tau_\tau}}
\rho(\frac{\sigma_{IP}}{c t},z) \frac{d\sigma_{IP}}{c t} dz
\nonumber\\
&=&
\left(
\sum_{n=1}^6  A_n\left(\frac{\sigma_{IP}}{c \tau_\tau}\right)^n
\right) \frac{d\sigma_{IP}}{\sigma_{IP}}dz,
\end{eqnarray}
where
\begin{eqnarray}
A_1&=&
        \sqrt{1/z-1}
         \left(\frac{12}5z+\frac{88}{15}z^2-\frac{64}{15}z^3
        +P_\tau(\frac{4}{15}z+\frac{32}{15}z^2-\frac{32}{5}z^3)\right)
        \e^{-\frac{\sigma_{IP}}{\tau_\tau\sqrt{1/z-1}}},
\nonumber\\
A_2&=& \left(-\frac{8}5z+\frac{58}{15}z^2-\frac{34}{15}z^3
        +P_\tau(\frac{4}{15}z+\frac{2}{15}z^2-\frac{2}{5}z^3)\right)
        \e^{-\frac{\sigma_{IP}}{\tau_\tau\sqrt{1/z-1}}}
\nonumber\\
&&      +\left(12 z^2-8z^3+P_\tau(4z^2-8z^3)\right)
         {\rm Ei}(-\frac{\sigma_{IP}}{\tau_\tau\sqrt{1/z-1}}),
\nonumber\\
A_3&=&
        \sqrt{1/z-1}
         \left(\frac{28}{15}z^2-\frac{38}{15}z^3
        +P_\tau(-\frac{8}{15}z^2-\frac{2}{15}z^3)\right)
        \e^{-\frac{\sigma_{IP}}{\tau_\tau\sqrt{1/z-1}}},
\nonumber\\
A_4&=& \left(\frac{1}{15}z^2-\frac{1}{15}z^3
        +P_\tau(-\frac{1}{15}z^2+\frac{1}{15}z^3)\right)
        \e^{-\frac{\sigma_{IP}}{\tau_\tau\sqrt{1/z-1}}} 
\nonumber\\
&&      +\left(2 z^2-\frac{8}{3}z^3-\frac{2}{3} P_\tau z^2 \right)
         {\rm Ei}(-\frac{\sigma_{IP}}{\tau_\tau\sqrt{1/z-1}}),
\nonumber\\
A_5&=&  
        \sqrt{1/z-1}
        \left(-\frac{1}{15}z^3
        +P_\tau \frac{1}{15}z^3\right)
        \e^{-\frac{\sigma_{IP}}{\tau_\tau\sqrt{1/z-1}}},
\nonumber\\
A_6&=& \left(-\frac{1}{15}z^3+\frac{1}{15} P_\tau z^3 \right)
        {\rm Ei}(-\frac{\sigma_{IP}}{\tau_\tau\sqrt{1/z-1}}).
\nonumber
\end{eqnarray}
Here $\tau_\tau$ is the life time of tau lepton, and ${\rm Ei}(x)$ is Exponential 
Integral,
\begin{eqnarray}
{\rm Ei}(-x) &=& -\int^\infty_x \frac1{t}\e^{-t} dt.
\nonumber
\end{eqnarray}

\newpage
%
%
\newcommand{\Journal}[4]{{\sl #1} {\bf #2} {(#3)} {#4}}
\newcommand{\APJ}{Ap. J.}
\newcommand{\CJP}{Can. J. Phys.}
\newcommand{\MPL}{Mod. Phys. Lett.}
\newcommand{\NC}{Nuovo Cimento}
\newcommand{\NP}{Nucl. Phys.}
\newcommand{\PL}{Phys. Lett.}
\newcommand{\PR}{Phys. Rev.}
\newcommand{\PRep}{Phys. Rep.}
\newcommand{\PRL}{Phys. Rev. Lett.}
\newcommand{\PTP}{Prog. Theor. Phys.}
\newcommand{\SJNP}{Sov. J. Nucl. Phys.}
\newcommand{\ZP}{Z. Phys.}

\newpage
\begin{table}[htbp]
  \begin{center}
    \leavevmode
    \begin{tabular}{|l|c|c|}
      \hline
      (GeV) & $\tan\beta=3$ & $\tan\beta=10$ \\
      \hline
      \hline
      $m_{\tilde{\chi}_1^0}$ &  56 & 58 \\
      \hline
      $m_{\tilde{\chi}_2^0}$ & 105 & 103 \\
      \hline
      $m_{\tilde{\chi}_1^-}$ & 100 & 100 \\
      \hline
      $m_{\tilde{\nu}_\mu}$ & 180 & 180 \\
        \hline
      $m_{\tilde{\mu}_L}$  & 194 & 197 \\
        \hline
      $m_{\tilde{\mu}_R}$ & 159 & 164 \\
        \hline
        $\mu$ & 244 & 200\\
      \hline
    \end{tabular}
    \caption{The sample SUSY parameters in our calculation.
        We fix the lightest chargino mass ($m_{\tilde{\chi}_1^-}$) 
        100 GeV and the muon sneutrino mass 
        ($ m_{\tilde{\nu}_\mu}$) 180 GeV. The lightest and the 
        second-lightest neutralino masses ($m_{\tilde{\chi}_1^0}$, 
        $m_{\tilde{\chi}_2^0}$),
        the left-handed and the right-handed smuon masses ($m_{\tilde{\mu}_L}$,
        $m_{\tilde{\mu}_R}$), and 
        other SUSY parameters are calculated by solving the RG 
        equations, assuming the minimal supergravity scenario and 
        the GUT relation for the gaugino masses. Here, 
        the Higgsino mass parameter $\mu$ is
        determined by the radiative electroweak symmetry
        breaking condition with $\mu>0$.}
    \label{tab:parameter}
  \end{center}
\end{table}
\begin{table}[htbp]
  \begin{center}
    \leavevmode
    \begin{tabular}{|c|c|c|}
      \hline
                        & $\tan\beta=3$ & $\tan\beta=10$ \\
      \hline
      \hline
      $\sel^+ \sel^-$     & 146         &  136           \\
      \hline
      $\sml^+ \sml^-$ ($\stl^+ \stl^-$)    & 30          &   26            \\
      \hline
      \hline
      $\sne \sne^c$     & 957         &  966           \\
      \hline
      $\snm \snm^c$($\snt \snt^c$)     & 18          &  18             \\
      \hline
    \end{tabular}
    \caption{Cross sections in fb 
for the left-handed slepton production at the $\e^+\e^-$
      collider with $\protect\sqrt{s}=500$ GeV. 
        We fix $m_{\snm}$=180 GeV, $\Delta m_{\tilde{\nu}}$=1 GeV and 
      $\theta_{\tilde{\nu}}$=0. The other SUSY parameters are the same
      as in Table~(1). 
      }
    \label{tab:lcross}
  \end{center}
\end{table}

\begin{table}[htbp]
  \begin{center}
    \leavevmode
    \begin{tabular}{|c|c|c|}
      \hline
                                        & $\tan\beta=3$ & $\tan\beta=10$ \\
      \hline
      \hline
      $Br(\sml^-(\sel^-) \to \mu^-(\e^-) \neu_{1}) $   & 0.05    & 0.08    \\
      \hline
      $Br(\sml^-(\sel^-) \to \mu^-(\e^-) \neu_{2}) $   & 0.39    & 0.41    \\
      \hline
      $Br(\sml^-(\sel^-) \to \nu_\mu(\nu_\e) \chb_{1}) $  & 0.56 & 0.51    \\
      \hline
      $Br(\stl^- \to \tau^- \neu_{1}) $  & 0.06   & 0.12    \\
      \hline
      $Br(\stl^- \to \tau^- \neu_{2}) $  & 0.38   & 0.38     \\
      \hline
      $Br(\stl^- \to \nu_\tau \chb_{1}) $ & 0.56  & 0.50      \\
      \hline
      \hline
      $Br(\snm(\sne) \to \nu_\mu(\nu_\e) \neu_{1}) $  & 0.30 &  0.26  \\
      \hline
      $Br(\snm(\sne) \to \nu_\mu(\nu_\e) \neu_{2}) $  & 0.14 &  0.15  \\
      \hline
      $Br(\snm(\sne) \to \mu^-(\e^-) \cha_{1}) $   & 0.56   &   0.59     \\
      \hline
      $Br(\snt \to \nu_\tau \neu_{1}) $ & 0.30 &  0.26   \\
      \hline
      $Br(\snt \to \nu_\tau \neu_{2}) $ & 0.14 &  0.15    \\
      \hline
      $Br(\snt \to \tau^- \cha_{1}) $  & 0.56    &   0.59     \\
      \hline
    \end{tabular}
    \caption{Branching ratios for the left-handed slepton decays.
     The input parameters are the same as in Table~(1).}
    \label{tab:ldecay}
  \end{center}
\end{table}

\begin{table}[htbp]
  \begin{center}
    \leavevmode
    \begin{tabular}{|l|c|c|}
      \hline
      & $\tan\beta=3$ & $\tan\beta=10$ \\
      \hline
      \hline
      ${Br}(\tilde{\chi}_1^-\rightarrow 2jets \tilde{\chi}_1^0)$ & 0.63 & 0.65\\
      \hline
      ${Br}(\tilde{\chi}_1^-\rightarrow l^-\bar{\nu}_l \tilde{\chi}_1^0) $ & 0.12$\times$3 &0.12$\times$3 \\
      \hline
      \hline
      ${Br}(\tilde{\chi}_2^0\rightarrow 2jets \tilde{\chi}_1^0)$ & 0.23 & 0.49 \\
      \hline
      ${Br}(\tilde{\chi}_2^0\rightarrow l^-  l^+ \tilde{\chi}_1^0)$ & 0.14$\times$3 & 0.09$\times$3\\
      \hline
      ${Br}(\tilde{\chi}_2^0\rightarrow \nu_l \bar{\nu}_l  \tilde{\chi}_1^0)$ & 0.12$\times$3 & 0.08$\times$3\\
      \hline
    \end{tabular}
    \caption{Branching ratios of the lightest chargino and 
      second-lightest neutralino three-body decays.
     The input parameters are the same as in Table~(1).}
    \label{tab:br}
  \end{center}
\end{table}

\begin{table}[htbp]
  \begin{center}
    \leavevmode
    \begin{tabular}{|c|c|c|}
      \hline
                        & $\tan\beta=3$ & $\tan\beta=10$ \\
      \hline
      \hline
      $\sml^+ \sml^-$     & 146           &  136          \\
      \hline
      $\stl^+ \stl^-$ ($\sel^+ \sel^-$ )    & 30            &  26          \\
      \hline
      \hline
      $\snm \snm^c$     & 957           &  966         \\
      \hline
      $\snt \snt^c$($\sne \sne^c$)     & 18            &   18         \\
      \hline
    \end{tabular}
    \caption{Cross sections in  fb 
for the left-handed slepton production at the $\mu^+\mu^-$
      collider with $\protect\sqrt{s}=500$ GeV. The input parameters are
      the same as in Table~(1).}
    \label{tab:lcrossm}
  \end{center}
\end{table}
\begin{table}
\begin{center}
\begin{tabular}{|c|c|c|c|}
\hline
$\sigma_{IP}^{\rm cut}$ &10$\mu m$&30$\mu m$&$(\infty)$\cr
\hline
$p_{\mu}$&0.02& 0.04 & 0.09
\cr
\hline 
$p_{\tau}$&0.88& 0.82&0.64
\cr
\hline
\end{tabular}
    \caption{The tau identification probability $p_{\tau}$ and 
        the the muon misidentification probability $p_{\mu}$
        for the IP cut $\sigma_{IP}^{\rm cut}=10$ and  $30\mu m$, and $\infty$.
        The parameter set is  the same as in Fig.~(6).
    }
    \label{tab:fmuftau}
\end{center}
\end{table}

\newpage
\clearpage
%
%
\begin{figure}[p]
\centerline{\psfig{file=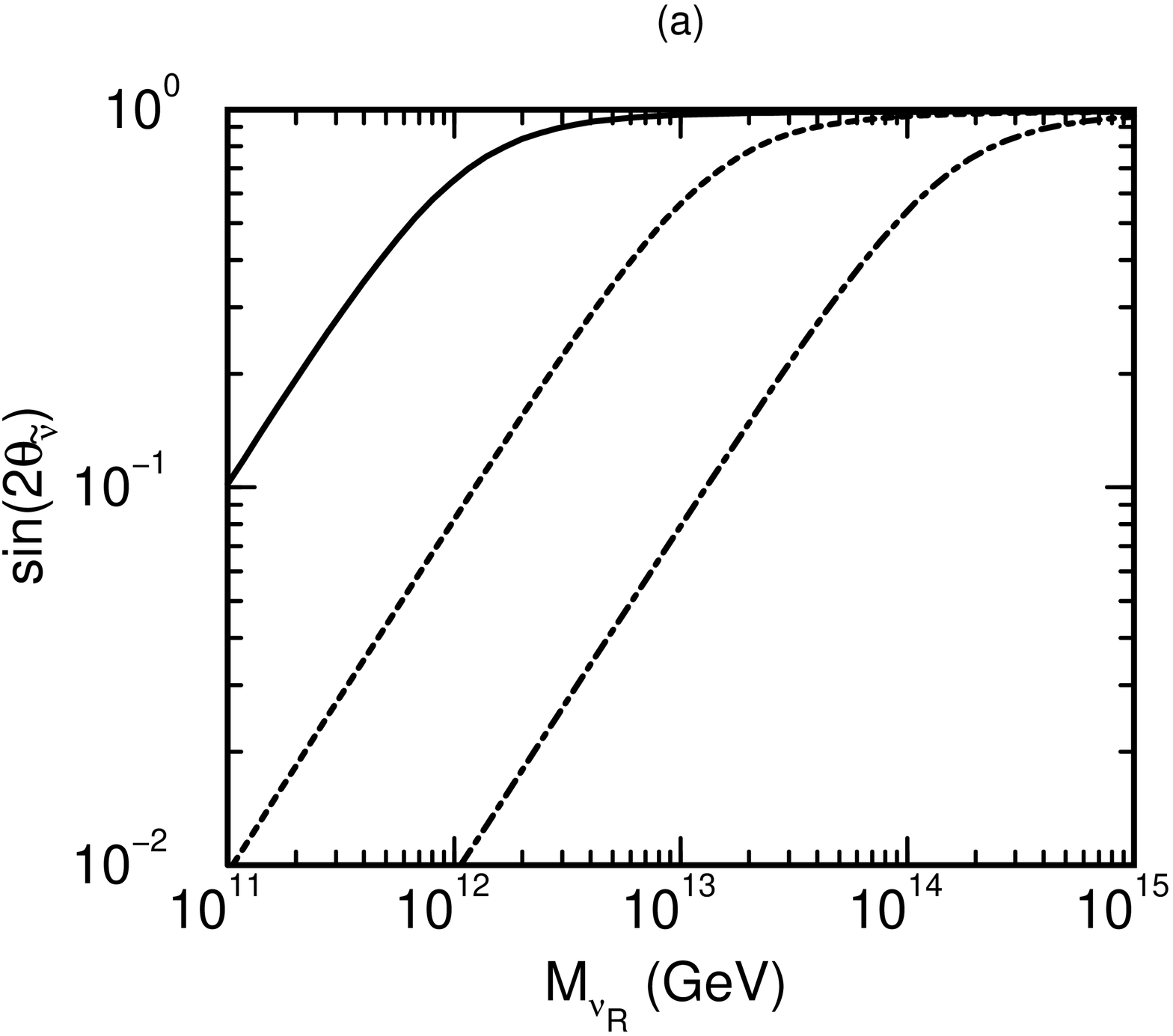,width=8cm,angle=-90}}
\vspace{1cm}

\centerline{\psfig{file=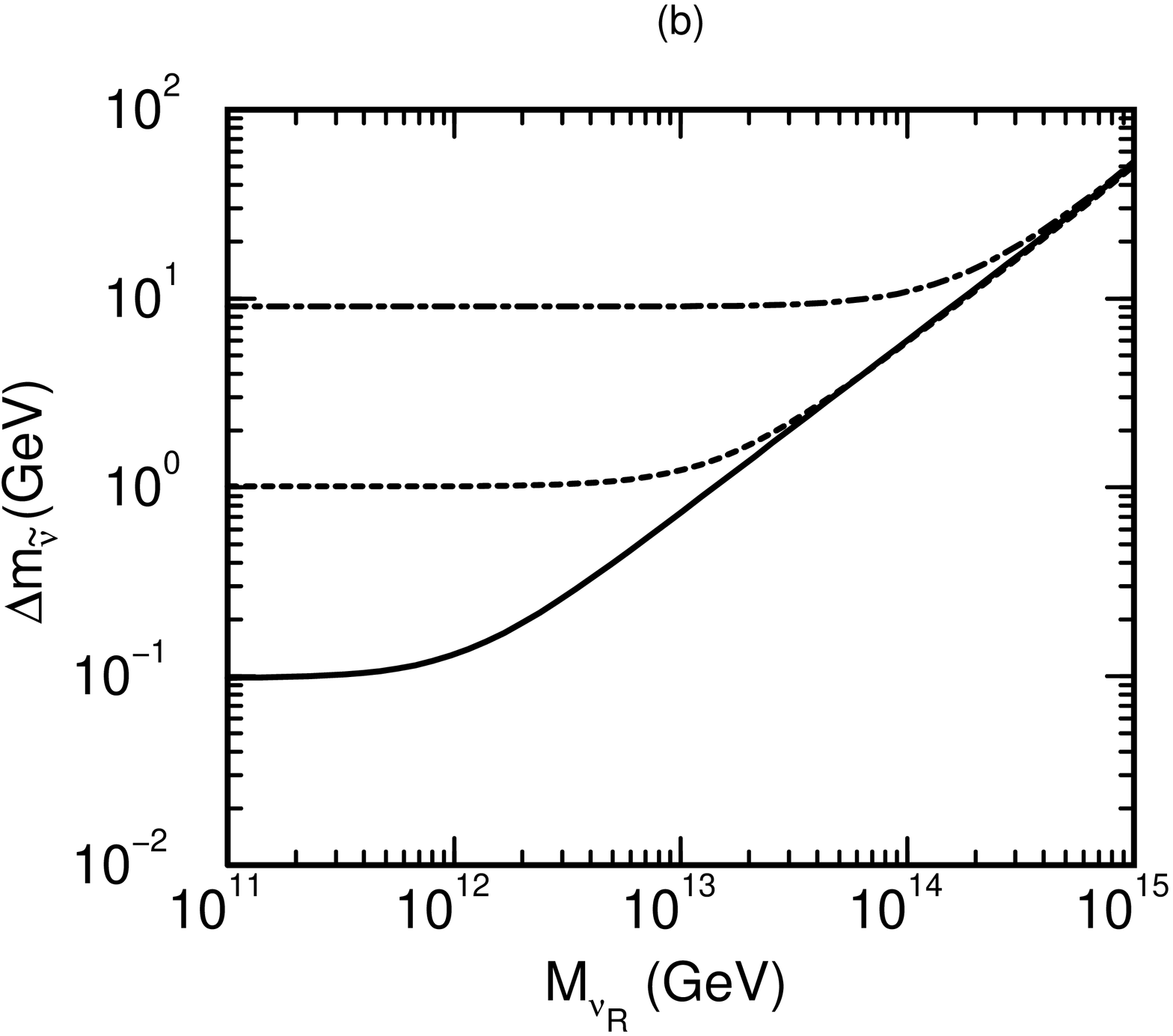,width=8cm,angle=-90}}
\caption{a) The mixing angle between the tau 
and muon sneutrinos ($\sin 2 \theta_{\tilde{\nu}}$) and b) the 
mass difference $(\Delta m_{\tilde{\nu}})$ between
the two non-$\e$-like sneutrinos 
as a function of the right-handed neutrino
scale. Here, we take $m^2_{\nu_\tau}=0.005\ {\rm eV}^2$, 
$\theta_D=\pi/4$, and $\bar{m}_{\tilde{\nu}}=180$ GeV. For simplicity, 
the gaugino masses in the MSSM are given by the GUT relation, and the 
lightest chargino mass is taken to be 100GeV. The other parameters
are determined in the minimal supergravity scenario and the radiative 
breaking condition of SU(2)$_L\times$U(1)$_Y$ with the Higgsino mass 
$(\mu)$ positive. Solid line, dashed line, and  dash-dot line 
are for $\tan\beta=3$, $10$, and $30$.}
\end{figure}
%
%
%
%
\begin{figure}[p]
\centerline{\psfig{file=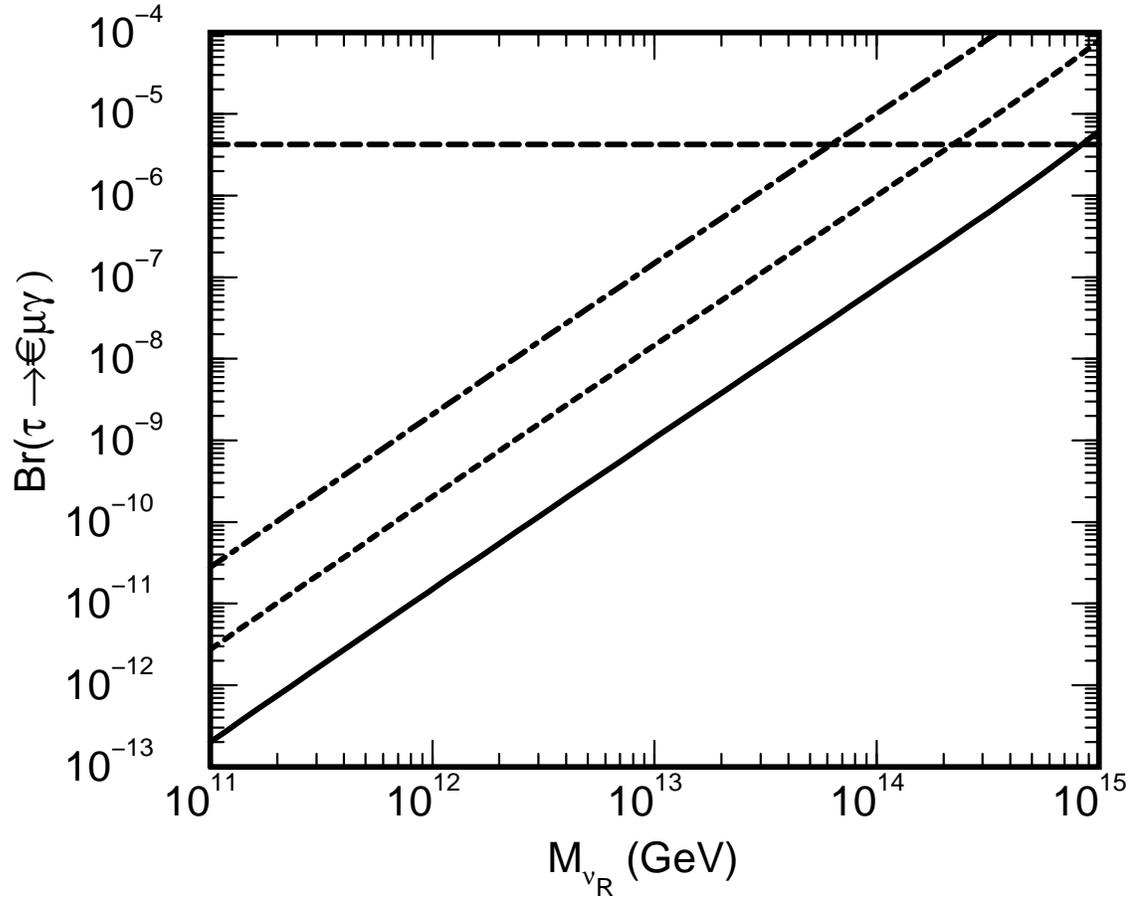,width=15cm,angle=-90}}
\caption{Branching ratio of $\tau\rightarrow \mu\gamma$ as a function 
of the right-handed neutrino scale. The input parameters are the same
as in Fig.~(1). The long-dashed line is the current experimental bound.
}
\end{figure}
%
%
%
%
%
%
\begin{figure}[p]
\centerline{\psfig{file=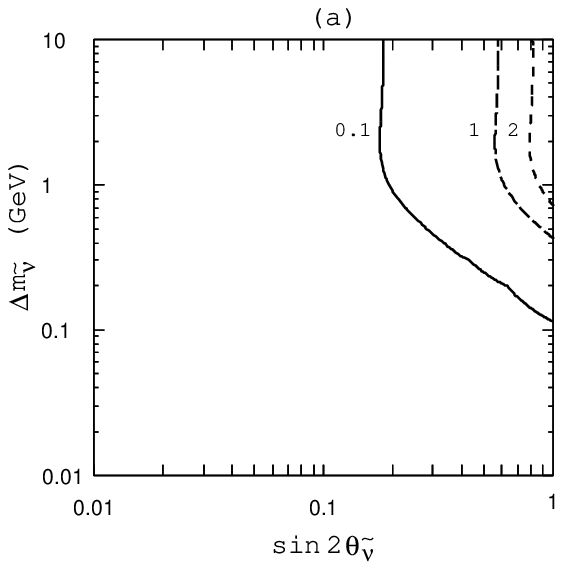,width=8cm}}

\vspace{1cm}
\centerline{\psfig{file=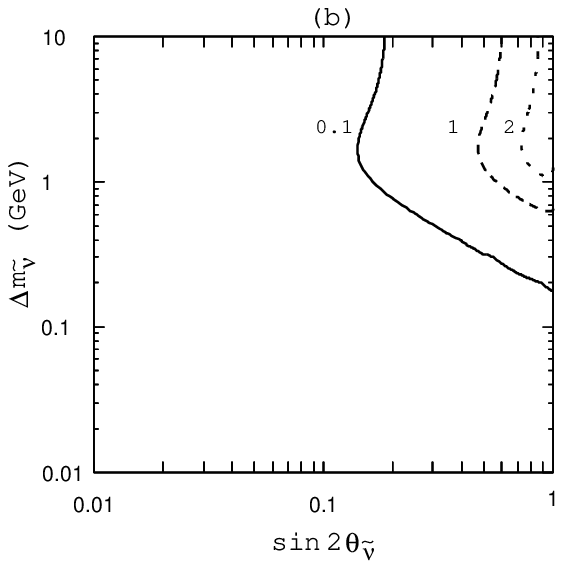,width=8cm}}
\caption{
Cross sections of (a) $\e^+\e^-\rightarrow\tau^+\mu^-  
+ \tilde{\chi}_1^- \tilde{\chi}_1^+$ and (b) $\e^+\e^-\rightarrow
\tau^+\mu^- +  \tilde{\chi}_2^0 \tilde{\chi}_2^0$ at 
the center mass energy 500GeV. We show them as
as functions of $\Delta m_{\tilde{\nu}}$ and  
$\sin 2 \theta_{\tilde{\nu}}$. 
Here, we take the sample SUSY parameter set for $\tan\beta=3$ listed 
in Table~(1).
}
\end{figure}
%
%
%
%
%
%
\begin{figure}[p]
\centerline{\psfig{file=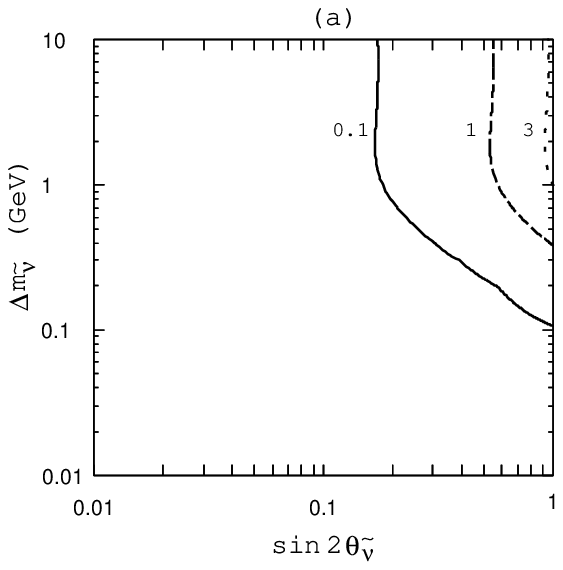,width=8cm}}

\vspace{1cm}
\centerline{\psfig{file=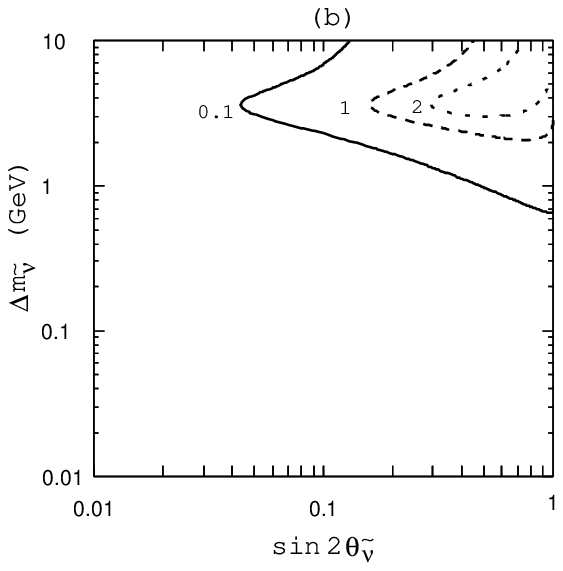,width=8cm}}
\caption{
Cross sections of (a) $\e^+\e^-\rightarrow\tau^+\mu^-  
+ \tilde{\chi}_1^- \tilde{\chi}_1^+$  and (b) $\e^+\e^-\rightarrow
\tau^+\mu^- +  \tilde{\chi}_2^0 \tilde{\chi}_2^0$  
at the center mass energy 500GeV.  We show them as functions of 
$\Delta m_{\tilde{\nu}}$ and $\sin 2 \theta_{\tilde{\nu}}$.
Here, we take the SUSY sample parameter set for $\tan\beta=10$ 
listed in Table~(1).}
\end{figure}
%
%
%
%
%
%
\begin{figure}[p]
\centerline{\psfig{file=fig5a.eps,width=8cm}}

\vspace{1cm}
\centerline{\psfig{file=fig5b.eps,width=8cm}}
\caption{
Cross sections of (a) $\mu^+\mu^-\rightarrow\tau^+\mu^-  
+ \tilde{\chi}_1^- \tilde{\chi}_1^+$ and (b) $\mu^+\mu^-\rightarrow
\tau^+\mu^- +  \tilde{\chi}_2^0 \tilde{\chi}_2^0$ 
at the center mass energy 500GeV. We show them as functions of 
$\Delta m_{\tilde{\nu}}$ and 
$\sin 2 \theta_{\tilde{\nu}}$. 
Here, we take the sample SUSY parameter set for $\tan\beta=3$ listed 
in Table~(1).
}
\end{figure}
%
%
%
%
%
%
\begin{figure}[p]
\centerline{\psfig{file=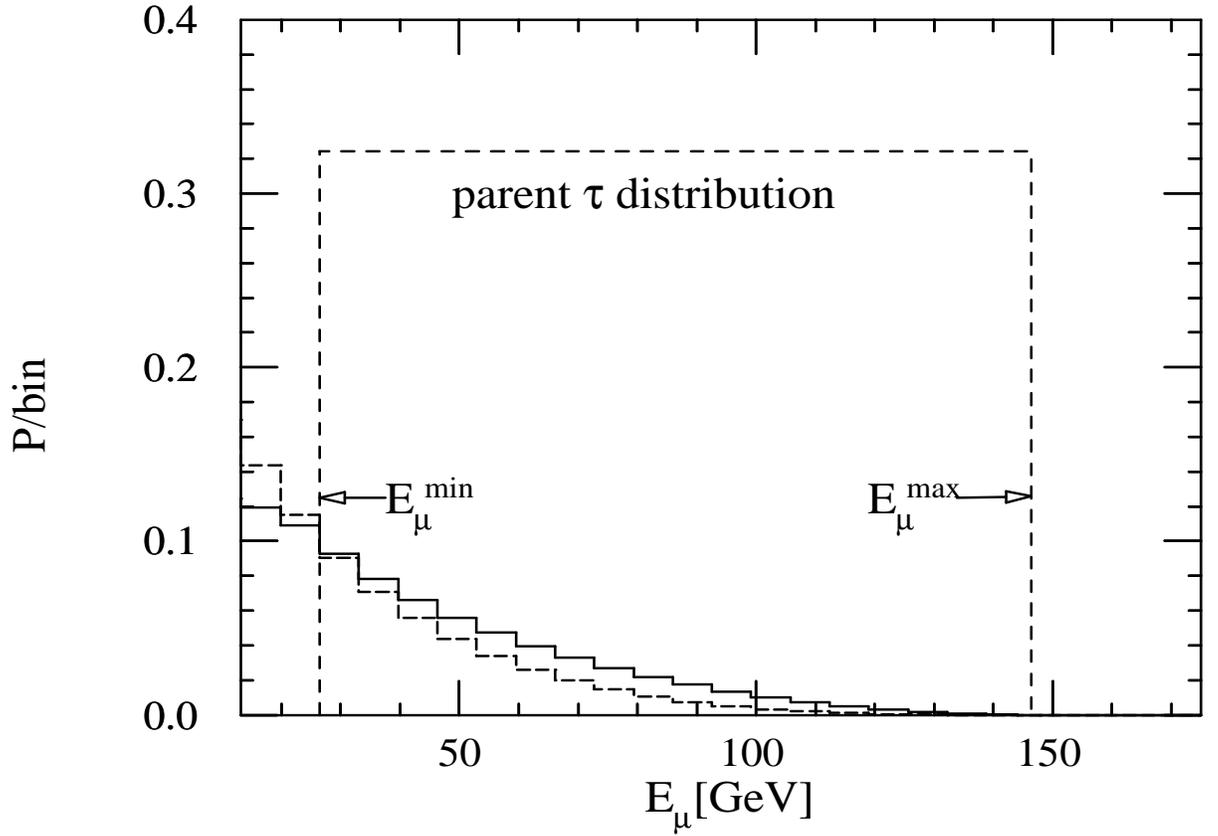,width=13cm,angle=-90}}
\caption{
Energy distribution of $\mu$ from decaying $\tau$ in flight.
The $\tau$ comes from the sneutrino pair production and the decay
to chargino with the center mass energy 
500GeV. Here, we take the sneutrino mass 180GeV, the chargino mass 100GeV. 
The energy distribution of the $\tau$ is flat between $E_\mu^{\rm min}$ 
and $E_\mu^{\rm max}$.
The solid and dashed lines are the energy distributions
of $\mu$ from decaying $\tau$ with polarization $-1$ and $+1$,
respectively. 
}
\end{figure}
%
%
%
%
%
%
\begin{figure}[p]
\centerline{\psfig{file=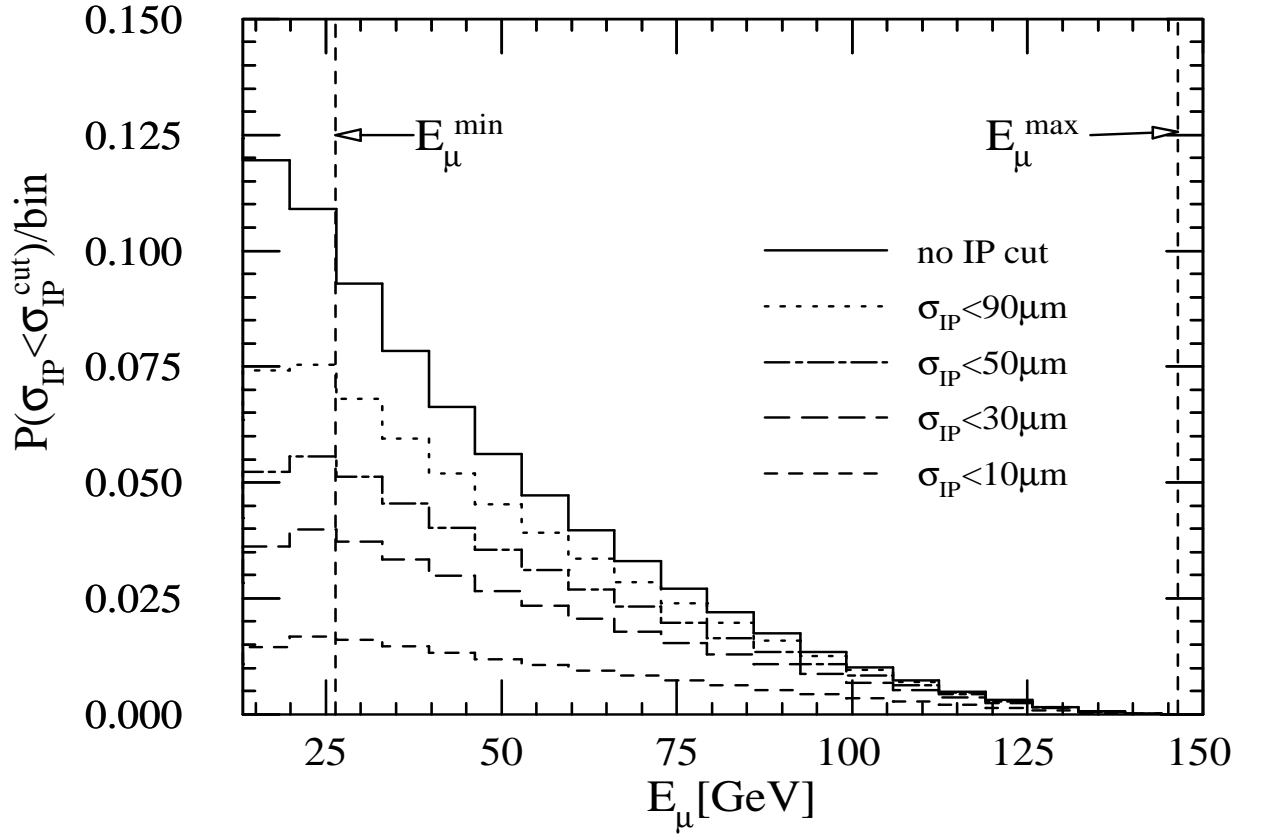,width=13cm,angle=-90}}
\caption{
Energy distribution of $\mu$ from decaying $\tau$, 
surviving after requiring the IP cut $\sigma_{IP}<\sigma^{\rm cut}_{IP}= 
10, 30, 50, 90\mu m$. The $\tau$ comes from the slepton decay.
The input parameter set is the same as in Fig.~(6).
}
\end{figure}
%
%
%
%
%
%
\begin{figure}[p]
\centerline{\psfig{file=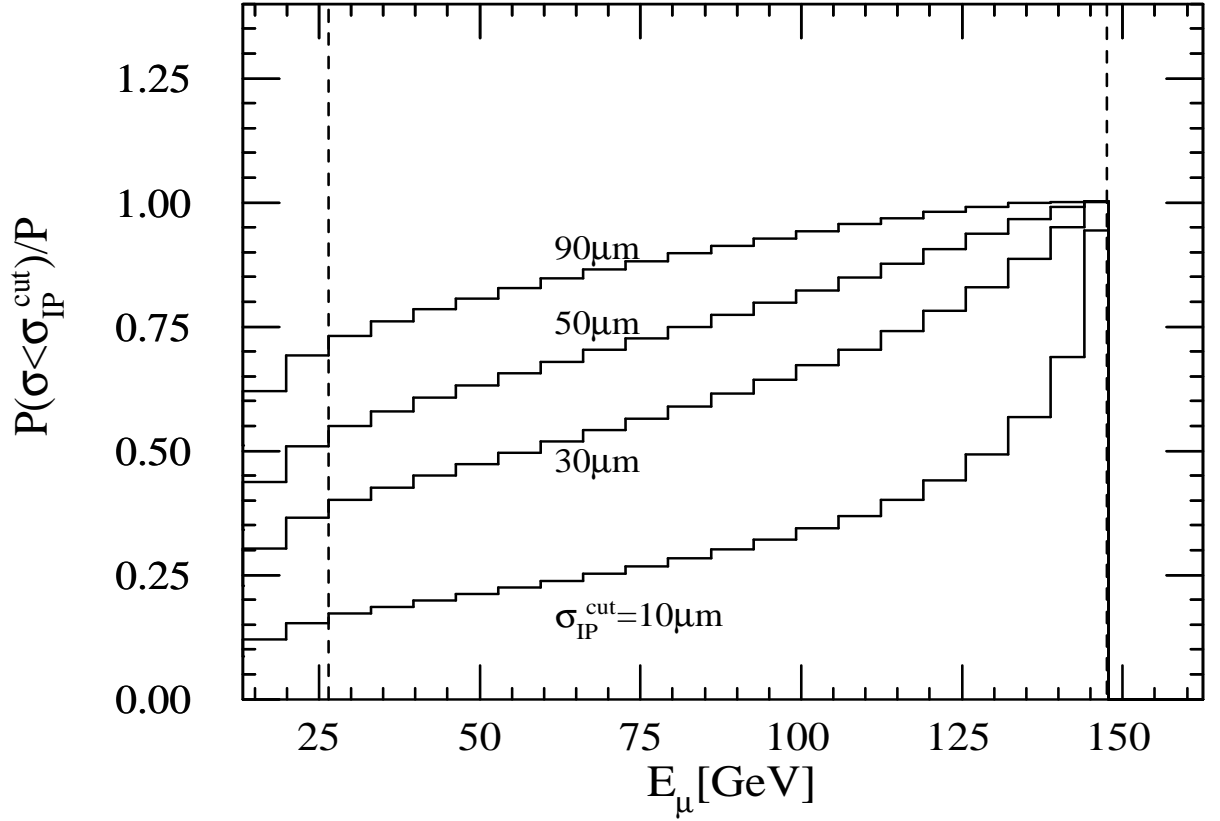,width=13cm,angle=-90}}
\caption{
Energy  distribution of $\mu$ from decaying $\tau$  after the IP cut
$\sigma_{IP}<\sigma_{IP}^{\rm cut}$ as the ratio to $E_{\mu}$ 
distribution without $\sigma_{IP}$ cut.
The input parameter set is the same as in Fig.~(6).
}
\end{figure}
%
%
%
%
%
%
\begin{figure}[p]
\centerline{\psfig{file=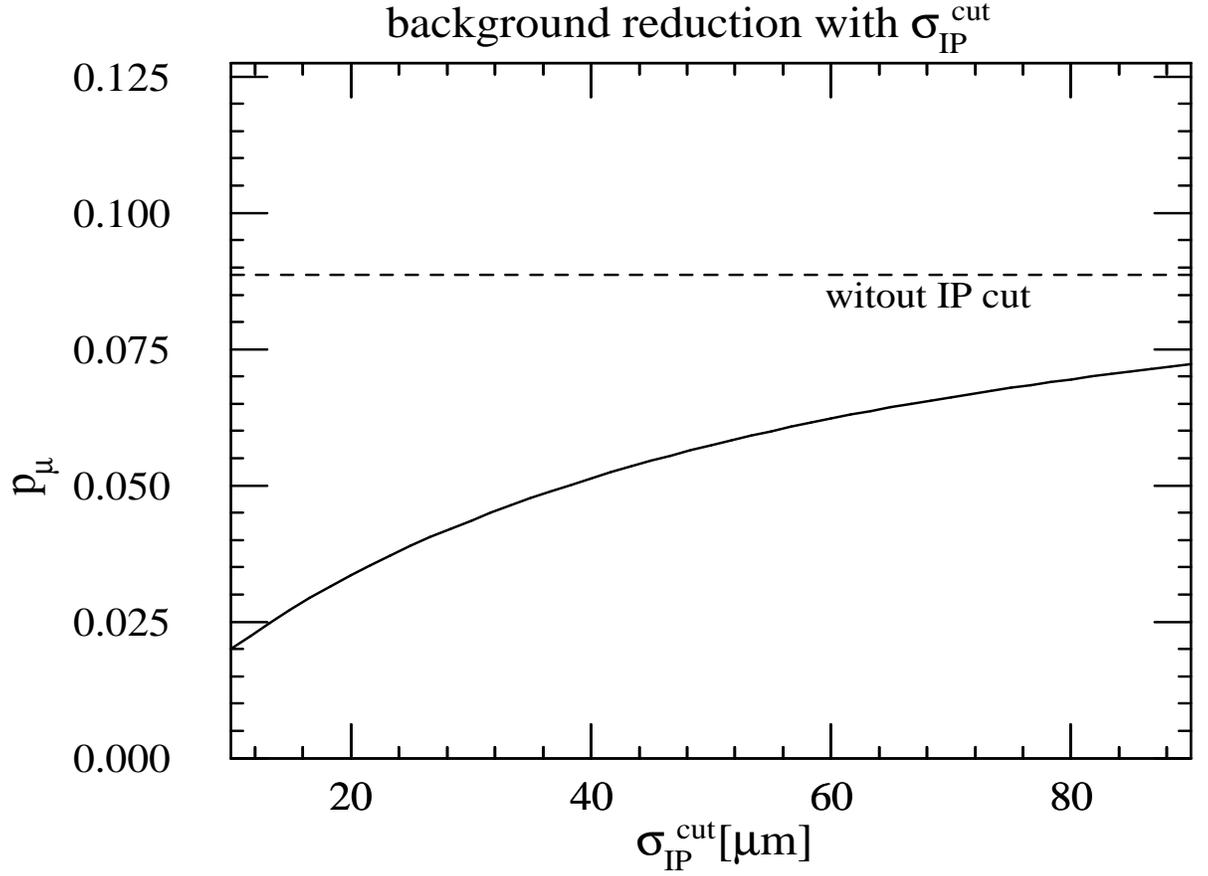,width=13cm,angle=-90}}
\caption{
Probability $p_\mu$ that 
$\tnu$ production and its decay 
into $\tau$ is mis-identified as $\tnu\rightarrow\mu\tchi$
when some  $\sigma^{\rm cut}_{IP}$ are applied in 
addition to the  energy cut $E_{\mu}>E^{\rm min}_{\mu}$. 
The parameter set is the same as in Fig.~(6).}
\end{figure}
%
%
%
%
%
\begin{figure}[p]
\centerline{\psfig{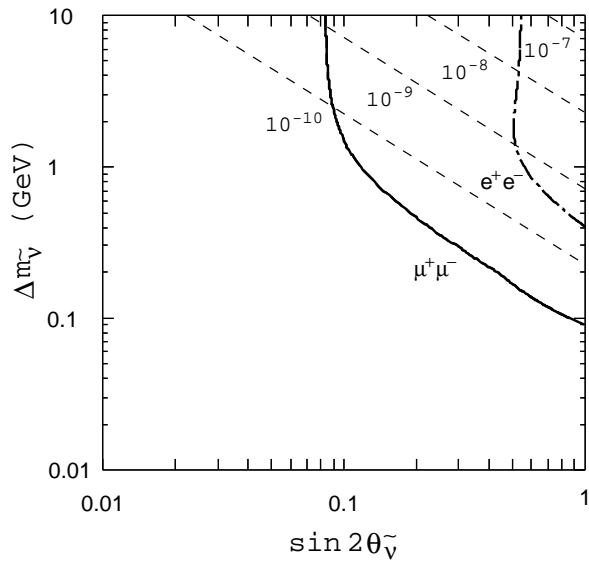}}
\caption{  
Significance contours corresponding to  
$3\sigma$ discovery as functions of 
$\sin2\theta_{\tilde{\nu}}$ 
and $\Delta m_{{\tilde{\nu}}}$. 
The dashed-dot (solid) line is for a $\e^+\e^-$ ($\mu^+\mu^-$) collider 
with the center mass energy 500GeV. 
We assume integrated luminosity ${\cal L}=50 {\rm fb}^{-1}$. 
For the $\e^+\e^-$ collider we take $\sigma^{\rm cut}_{IP}= 10 \mu m$. 
Here, we take the sample SUSY parameter 
set for $\tan\beta=3$ listed in Table~(1). 
We also show contours of the constant $\tau \rightarrow \mu \gamma$ branching 
ratio, $10^{-7}$, $10^{-8}$, $10^{-9}$, and $10^{-10}$ by the dashed lines.
}
\end{figure}

\end{document}